\documentclass[twocolumn, floatfix, tighten, times, trackchanges]{aastex631}
\usepackage[utf8]{inputenc}
\usepackage{amstext}

\usepackage{graphicx}
\usepackage{xspace}
\usepackage{acronym}
\usepackage{longtable}
\AtBeginEnvironment{longtable}{\tiny}
\definecolor{lightgray}{gray}{0.9}
\usepackage{booktabs}
\usepackage[caption=false]{subfig}
\usepackage{siunitx}

\newcommand\gwtc[1][?]{\mbox{GWTC\if#1?\else-#1\fi}}
\newcommand\thisgwtcversionmajor{5}
\newcommand\thisgwtcversionminor{0}
\newcommand\thisgwtcversionfull{\thisgwtcversionmajor.\thisgwtcversionminor}
\newcommand\thisgwtcversion\thisgwtcversionfull

\providecolor{UNREVIEWED}{rgb}{1,0,0}
\providecolor{REVIEWED}{rgb}{0,.5,.5}
\providecolor{NOTAPPLICABLE}{rgb}{0,0.5,0}

\newcommand{\soft}[1]{\textsc{#1}}

\newcommand{\BILBY}{\soft{Bilby}\xspace}

\newcommand{\GRACEDB}{\soft{GraceDB}\xspace}

\providecolor{UNREVIEWED}{rgb}{1,0,0}
\providecolor{REVIEWED}{rgb}{0,.5,.5}
\providecolor{NOTAPPLICABLE}{rgb}{0,0.5,0}

\DeclareSIUnit\parsec{pc}
\DeclareSIUnit\Mpc{\mega\parsec}
\DeclareSIUnit\yr{yr}
\DeclareSIUnit\GpcCubedYear{\giga\parsec\cubed\yr}

\newcommand{\Msun}{\ensuremath{\mathit{M_\odot}}}

\newcommand{\Mc}{\ensuremath{\mathcal{M}}}
\newcommand{\Mtot}{\ensuremath{M}}

\newcommand{\chieff}{\ensuremath{\chi_\mathrm{eff}}}
\newcommand{\chip}{\ensuremath{\chi_\mathrm{p}}}

\newcommand{\DL}{\ensuremath{D_\mathrm{L}}}

\newcommand\PEpdfp{\ensuremath{p}}

\newcommand{\PEparameter}{\ensuremath{\boldsymbol{\theta}}}%

\newcommand\PEpdf[2][?]{\ensuremath{\PEpdfp({#2}\ifx#1?\else | {#1}\fi)}} 

\newcommand\PEpriorpdfpi{\ensuremath{\pi}} 
\newcommand\PEpdfprior[1]{\ensuremath{\PEpriorpdfpi({#1})}} 
\newcommand\PEprior[1][\PEparameter]{\PEpdfprior{#1}} 
\newcommand\PEpriorpe[1][\PEparameter]{{\let\keepPEpriorpdfpi\PEpriorpdfpi\def\PEpriorpdfpi{\keepPEpriorpdfpi_{\text{PE}}}\PEprior[#1]\let\PEpriorpdfpi\keepPEpriorpdfpi}} 

\newcommand{\SNR}[0]{\ac{SNR}\xspace}

\usepackage{amstext}

\makeatletter
\@ifpackageloaded{cleveref}{
  \AtBeginDocument{
    \def\ltx@label#1{\cref@label{#1}}
    \def\label@in@display@noarg#1{\cref@old@label@in@display{#1}}
    \def\label@in@mmeasure@noarg#1{%
      \begingroup%
        \measuring@false%
        \cref@old@label@in@display{#1}
      \endgroup}%
  }
}{}
\makeatother

\protected\def\protectedacused{\acused}

\acrodef{LIGO}[LIGO]{Laser Interferometer Gravitational-Wave Observatory}
\acrodef{LHO}[LHO]{\ac{LIGO} Hanford Observatory}
\acrodef{LLO}[LLO]{\ac{LIGO} Livingston Observatory}
\acrodef{KAGRA}[KAGRA]{KAGRA}\acused{KAGRA}
\acrodef{iKAGRA}[iKAGRA]{initial \ac{KAGRA}}
\acrodef{bKAGRA}[bKAGRA]{baseline \ac{KAGRA}}
\acrodef{GEO}[GEO\,600]{GEO\,600 \ac{GW} detector}
\acrodef{aLIGO}{Advanced \ac{LIGO}}
\acrodef{A+}{Advanced+ \ac{LIGO}}
\acrodef{Asharp}[\ensuremath{\text{A}^\sharp}]{\ac{LIGO} \acs{Asharp}}
\acrodef{AdV}{Advanced \acl{Virgo}}
\acrodef{AdV+}{Advanced \acl{Virgo}+}
\acrodef{Virgo}{Virgo}\acused{Virgo}
\acrodef{VirgoNEXT}[Virgo\_nEXT]{Virgo\_nEXT}\acused{VirgoNEXT}

\acrodef{LSC}[LSC]{\acs{LIGO} Scientific Collaboration}
\acrodef{LV}[LV]{\acs{LIGO}--\acs{Virgo} Collaboration\protect\protectedacused{LVC}}
\acrodef{LVC}[LV]{\acs{LIGO}--\acs{Virgo} Collaboration\protect\protectedacused{LV}}
\acrodef{LVK}[LVK]{\acs{LIGO}--\ac{Virgo}--\ac{KAGRA} Collaboration}
\acrodef{IGWN}[IGWN]{International \ac{GWH} Observatory Network}

\acrodef{O1}[O1]{first observing run}
\acrodef{O2}[O2]{second observing run}
\acrodef{O3}[O3]{third observing run}
\acrodef{O3a}[O3a]{first half of the third observing run}
\acrodef{O3b}[O3b]{second half of the third observing run}
\acrodef{O3GK}[O3GK]{observing run}
\acrodef{O4}[O4]{fourth observing run}
\acrodef{O4a}[O4a]{first part of the fourth observing run}
\acrodef{O4b}[O4b]{second part of the fourth observing run}
\acrodef{O4c}[O4c]{third part of the fourth observing run}
\acrodef{IR1}[IR1]{intermediate run 1}
\acrodef{O5}[O5]{fifth observing run}

\acrodef{BH}[BH]{black hole}
\acrodef{BBH}[BBH]{binary \ac{BH}}
\acrodef{BNS}[BNS]{binary \ac{NS}}
\acrodef{IMBH}[IMBH]{intermediate-mass \ac{BH}}
\acrodef{NS}[NS]{neutron star}
\acrodef{BHNS}[BHNS]{\ac{BH}--\ac{NS} binary}
\acrodefplural{BHNS}[BHNSs]{\ac{BH}--\ac{NS} binaries}
\acrodef{NSBH}[NSBH]{\ac{NS}--\ac{BH} binary}
\acrodefplural{NSBH}[NSBHs]{\ac{NS}--\ac{BH} binaries}
\acrodef{PBH}[PBH]{primordial \ac{BH}}
\acrodef{CBC}[CBC]{compact binary coalescence}

\acrodef{GW}[GW]{gravitational wave\protect\protectedacused{GWH}}
\acrodef{GWH}[GW]{gravitational-wave\protect\protectedacused{GW}}
\acrodef{IFO}[IFO]{interferometer}
\acrodef{SNR}[SNR]{signal-to-noise ratio}
\acrodef{FAR}[FAR]{false-alarm rate}
\acrodef{IFAR}[IFAR]{inverse false-alarm rate}
\acrodef{FAP}[FAP]{false alarm probability}
\acrodef{PSD}[PSD]{power spectral density}
\acrodef{GR}[GR]{general relativity}
\acrodef{NR}[NR]{numerical relativity}
\acrodef{PN}[PN]{post-Newtonian}
\acrodef{EOB}[EOB]{effective-one-body}
\acrodef{ROM}[ROM]{reduced-order model}
\acrodef{IMR}[IMR]{inspiral--merger--ringdown}
\acrodef{PDF}[pdf]{probability density function}
\acrodef{PE}[PE]{parameter estimation}
\acrodef{CI}[CI]{credible interval}
\acrodef{CL}[CL]{credible level}
\acrodef{EOS}[EoS]{equation of state}
\acrodef{KLD}[KLD]{Kullback--Leibler divergence}
\acrodef{JSD}[JSD]{Jensen--Shannon divergence}
\acrodef{GCN}[GCN]{General Coordinates Network}
\acrodef{GWTC}[GWTC]{Gravitational-Wave Transient Catalog}
\acrodef{GWOSC}[GWOSC]{Gravitational Wave Open Science Center}
\acrodef{WDM}[WDM]{Wilson--Daubechies--Meyer}
\acrodef{DQR}[DQR]{data-quality report}

\acrodef{CWB}[cWB]{coherent WaveBurst}
\acrodef{LAL}[LAL]{\ac{LIGO} algorithm library}

\acrodef{CHRoCC}{central heating radius of curvature correction}
\acrodef{NonSENS}{non-stationary estimation and noise subtraction}
\acrodef{RF}{radio frequency}
\acrodef{PNC}{phase noise cancellation}
\acrodef{ASC}{alignment sensing and control}
\acrodef{WFS}{wave-front sensing}
\acrodef{BPC}{beam position control}
\acrodef{ADS}{alignment dither systems}
\acrodef{OMC}{output mode cleaner}
\acrodef{LVDTs}{linear variable differential transformers}
\acrodef{GAS}{geometrical anti-spring}

\acrodef{PTA}{Pulsar Timing Array}

\newcolumntype{Y}{>{\centering\arraybackslash}X}
\newcolumntype{Z}{>{\raggedright\arraybackslash}X}
\newcolumntype{K}{>{\raggedleft\arraybackslash}X}
\newcolumntype{U}{>{\hsize=1.01\hsize}Y}
\newcolumntype{V}{>{\hsize=1.2\hsize}Y}
\newcolumntype{W}{>{\hsize=0.71\hsize}Y}

\usepackage{scrextend}
\usepackage{textgreek}
\usepackage{graphicx}
\usepackage[caption=false]{subfig}

\usepackage{makerobust}

\renewcommand{\today}{\number\day\space\ifcase\month\or
  January\or February\or March\or April\or May\or June\or
  July\or August\or September\or October\or November\or December\fi
  \space\number\year}

\newcommand{\networkTWOONEmultipolesnrXPHM}{\ensuremath{1.3^{+1.4}_{-1.0}}}
\newcommand{\networkTHREETHREEmultipolesnrXPHM}{\ensuremath{2.2^{+1.6}_{-1.6}}}
\newcommand{\networkFOURFOURmultipolesnrXPHM}{\ensuremath{3.3^{+0.5}_{-1.7}}}

\newcommand{\LONEoptimalsnrcombined}{\ensuremath{42.0^{+1.6}_{-1.7}}}

\newcommand{\chieffcombined}{\ensuremath{-0.01^{+0.06}_{-0.07}}}

\newcommand{\chirpmasssourcecombined}{\ensuremath{26.8^{+0.9}_{-1.0}}}

\newcommand{\costhetajncombined}{\ensuremath{0.0^{+0.7}_{-0.7}}}

\newcommand{\finalmasscombined}{\ensuremath{62.7^{+1.5}_{-1.4}}}

\newcommand{\finalspincombined}{\ensuremath{0.68^{+0.02}_{-0.02}}}

\newcommand{\luminositydistancecombined}{\ensuremath{289^{+183}_{-130}}}

\newcommand{\massONEsourcecombined}{\ensuremath{33.7^{+2.9}_{-2.2}}}

\newcommand{\massTWOsourcecombined}{\ensuremath{28.2^{+2.2}_{-3.1}}}
\newcommand{\massratiocombined}{\ensuremath{0.84^{+0.12}_{-0.14}}}

\newcommand{\networkmatchedfiltersnrcombined}{\ensuremath{42.13^{+0.07}_{-0.11}}}
\newcommand{\networkoptimalsnrcombined}{\ensuremath{42.0^{+1.6}_{-1.7}}}

\newcommand{\redshiftcombined}{\ensuremath{0.06^{+0.04}_{-0.03}}}

\newcommand{\spinONEzcombined}{\ensuremath{0.0^{+0.1}_{-0.2}}}

\newcommand{\spinTWOzcombined}{\ensuremath{0.0^{+0.3}_{-0.1}}}

\newcommand{\totalmasssourcecombined}{\ensuremath{61.9^{+2.0}_{-2.0}}}

\newcommand{\chipcombinedzerobound}{\ensuremath{0.42}}

\newcommand{\aONEmagcombinedzerobound}{\ensuremath{0.38}}
\newcommand{\aTWOmagcombinedzerobound}{\ensuremath{0.50}}

\newcommand{\reviewed}[1]{#1}

\newcommand{\Mfdet}{\ensuremath{M_\mathrm{f}^{\mathrm{det}}}}

\newcommand{\UTCTIME}{\reviewed{2023-08-14 23:09:01.805 UTC}}
\newcommand{\OPTIMALSNR}{\reviewed{\LONEoptimalsnrcombined}}

\newcommand{\SNRgstlalonlinerounded}{\reviewed{42.4}}
\newcommand{\SNRTWOONE}{\reviewed{\networkTWOONEmultipolesnrXPHM}}
\newcommand{\SNRTHREETHREE}{\reviewed{\networkTHREETHREEmultipolesnrXPHM}}
\newcommand{\SNRFOURFOUR}{\reviewed{\networkFOURFOURmultipolesnrXPHM}}
\newcommand{\IMRCTCUT}{\reviewed{144}}
\newcommand{\IMRCTLOWSNR}{\reviewed{35.3}}
\newcommand{\IMRCTHIGHSNR}{\reviewed{21.7}}
\newcommand{\RESIDUALP}{\reviewed{0.85}}

\definecolor{notecolor}{rgb}{0.0, 0.1, 1.0}

\newcommand{\dphi}[1]{\ensuremath{\delta\hat{\varphi}_{#1}}}

\newacro{gwtc4}[GWTC-4.0]{the fourth Gravitational-Wave Transient Catalog}
\newacro{gwtc3}[GWTC-3.0]{the third Gravitational-Wave Transient Catalog}
\newacro{lvk}[LVK]{LIGO--Virgo--KAGRA collaboration}
\acrodef{LLO}[LLO]{LIGO Livingston observatory}
\newacro{seobnr}[SEOBNR]{\textsc{SEOBNRv5PHM}}
\newacro{seobnrhm}[v5HM]{\textsc{SEOBNRv5HM}}
\newacro{seobnrhmrom}[SEOBNR]{\textsc{SEOBNRv5HM$\_$ROM}}
\newacro{xphm}[XPHM]{\textsc{IMRPhenomXPHM$\_$SpinTaylor}}
\newacro{xo4a}[XO4a]{\textsc{IMRPhenomXO4a}}
\newacro{nrsur}[NRSur]{\textsc{NRSur7dq4}}
\newacro{nr}[NR]{numerical-relativity}
\newacro{qnm}[QNM]{quasi-normal mode}
\newacro{IMRCT}[IMRCT]{Inspiral\textendash Merger\textendash Ringdown Consistency Test}
\newacro{qnmrf}[QNMRF]{Quasi-Normal Mode Rational Filter}
\newacro{pseobnr}[pSEOBNR]{pSEOBNRv5PHM}
\newacro{maxL}[maxL]{maximum-likelihood}
\newacro{ds}[DS]{damped-sinusoid}

\begin{document}

\title{GW230814: investigation of a loud gravitational-wave signal observed with a single detector}

\author[0000-0003-4786-2698]{A.~G.~Abac}
\affiliation{Max Planck Institute for Gravitational Physics (Albert Einstein Institute), D-14476 Potsdam, Germany}
\author{I.~Abouelfettouh}
\affiliation{LIGO Hanford Observatory, Richland, WA 99352, USA}
\author{F.~Acernese}
\affiliation{Dipartimento di Farmacia, Universit\`a di Salerno, I-84084 Fisciano, Salerno, Italy}
\affiliation{INFN, Sezione di Napoli, I-80126 Napoli, Italy}
\author[0000-0002-8648-0767]{K.~Ackley}
\affiliation{University of Warwick, Coventry CV4 7AL, United Kingdom}
\author[0000-0001-5525-6255]{C.~Adamcewicz}
\affiliation{OzGrav, School of Physics \& Astronomy, Monash University, Clayton 3800, Victoria, Australia}
\author[0009-0004-2101-5428]{S.~Adhicary}
\affiliation{The Pennsylvania State University, University Park, PA 16802, USA}
\author{D.~Adhikari}
\affiliation{Max Planck Institute for Gravitational Physics (Albert Einstein Institute), D-30167 Hannover, Germany}
\affiliation{Leibniz Universit\"{a}t Hannover, D-30167 Hannover, Germany}
\author[0000-0002-4559-8427]{N.~Adhikari}
\affiliation{University of Wisconsin-Milwaukee, Milwaukee, WI 53201, USA}
\author[0000-0002-5731-5076]{R.~X.~Adhikari}
\affiliation{LIGO Laboratory, California Institute of Technology, Pasadena, CA 91125, USA}
\author{V.~K.~Adkins}
\affiliation{Louisiana State University, Baton Rouge, LA 70803, USA}
% ...

\collaboration{3000}{The LIGO Scientific Collaboration, the Virgo Collaboration, and the KAGRA Collaboration}


\begin{abstract}

GW230814, detected by the LIGO Livingston observatory with a signal-to-noise ratio of \SNRgstlalonlinerounded\, represents the loudest gravitational-wave signal in the GWTC-4.0 catalog.
Its source is consistent with a binary black hole coalescence with
component masses $m_1 = \massONEsourcecombined\,\Msun$, $m_2=\massTWOsourcecombined\,\Msun$,
and a small effective inspiral spin \reviewed{$\chi_{\rm eff} = \chieffcombined$}.
The high signal-to-noise ratio enabled us to detect an 
$\ell=|m|=4$ mode in the inspiral-merger-ringdown signal for the first time (with \(\log_{10} \mathrm{BF}\simeq 1 \)),
and enables a range of tests of consistency between theoretical predictions and the observed waveform.
While most of these tests show agreement with theoretical predictions, there are suggestions of minor deviations in the ringdown phase.
Simulations that incorporate general relativity and realistic detector noise reproduce similar deviations, suggesting they do not constitute evidence for a breakdown of general relativity.
The observation of GW230814 demonstrates that the unprecedented sensitivity of
the detectors enables highly significant detections with a single observatory.
However, without corroborating data from a multi-detector network, the ability to draw rigorous conclusions about fundamental physics remains severely limited.

\end{abstract}

\pacs{%
04.80.Nn, 
04.25.dg, 
95.85.Sz, 
97.80.-d   
04.30.Db, 
04.30.Tv  
}

\section{Introduction}\label{sec:intro}

\begin{figure*}
    \centering
    \includegraphics[width=0.98\textwidth]{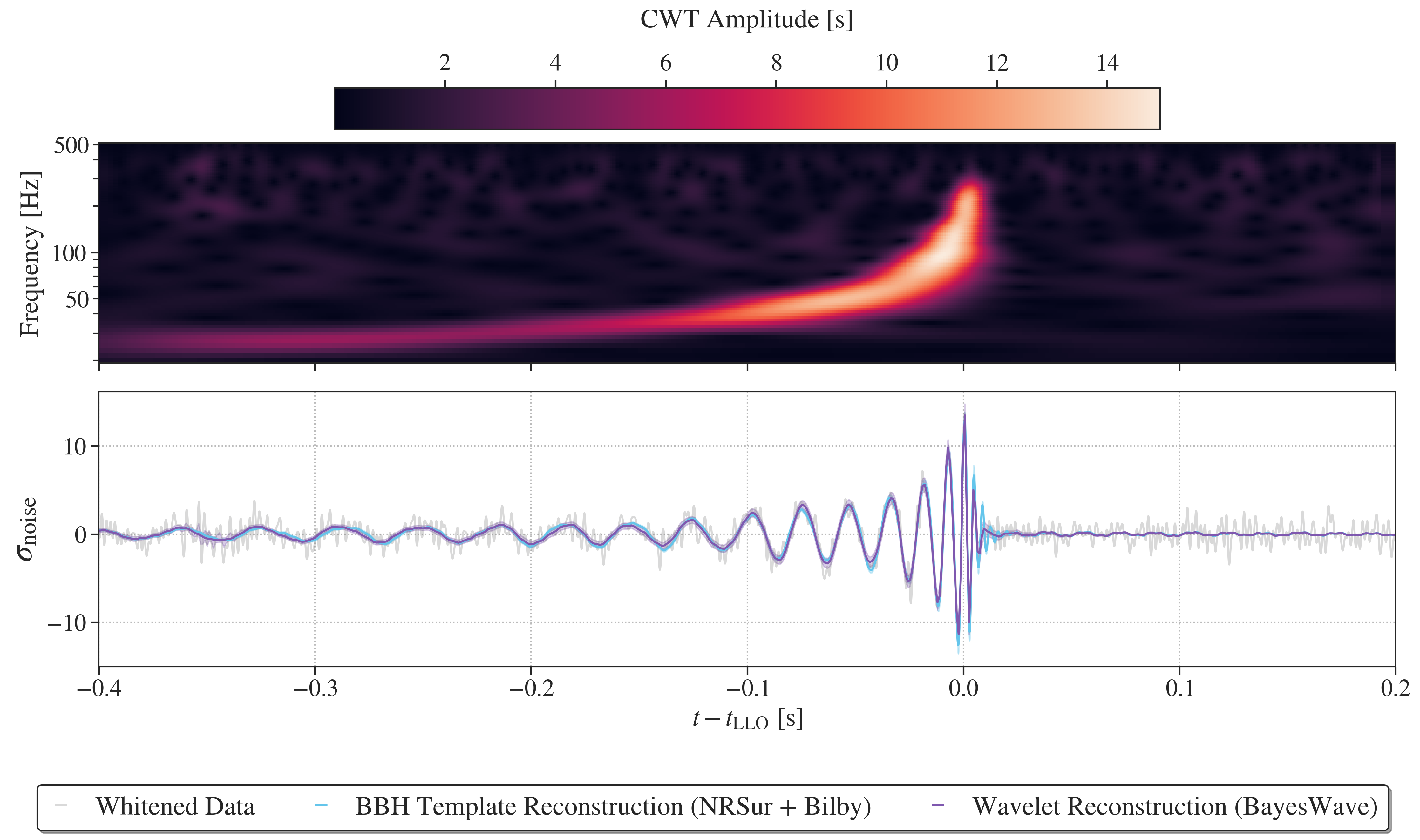}
    \caption{
    Time-frequency representation (top) and whitened time-domain strain with Bayesian reconstructions (bottom) from LIGO Livingston observations of GW230814.  
Time is measured relative to GPS time \( t_{\rm LLO} = \reviewed{\input{generated-values/L1-peaktime.tex}} \)\,s.  
The top panel shows the time-frequency representation of the whitened strain data, obtained using a continuous wavelet transform (CWT) with a Morlet–Gabor wavelet. The color scale represents the amplitude of the CWT coefficients, in units of seconds.  
The bottom panel displays the whitened strain data (gray) alongside 90\% credible intervals from waveform reconstructions using the \acf{nrsur} model (blue) and the wavelet-based BayesWave algorithm (purple).  
The data are high-pass filtered at 20\,Hz and downsampled to 1024\,Hz in both panels.
    }
    \label{fig:data}
\end{figure*}

The \ac{GW} signal GW230814\_230901, henceforth referred to as GW230814, was observed by the \ac{LLO} at \UTCTIME. 
At the time of the observation, only \ac{LLO} was taking data, with the LIGO Hanford, Virgo and KAGRA detectors offline.
Despite this, the signal, with an inferred optimal \SNR of \OPTIMALSNR, is louder than any \ac{CBC} published by the \ac{lvk} in \ac{gwtc4}~\citep{LIGOScientific:2025slb} or in non-\ac{lvk} catalogs~\citep{2023ApJ...946...59N, Wadekar:2023gea}. The Livingston data, shown in Fig.~\ref{fig:data}, clearly shows the presence of a loud \ac{BBH} coalescence signal.
Due to its high \ac{SNR}, the signal is of particular interest for its ability to constrain the properties of the source, including measurement of the inclination purely through the relative amplitudes of the different multipole modes of the radiation. In this work we expand upon the analysis presented in ~\citet{LIGOScientific:2025slb}.

The signal was detected in low latency by the GstLAL pipeline~\citep{sakon2023template, ewing2023performance, 2023PhRvD.108d3004T} with an \ac{IFAR} greater than $10^5$ years, and later confirmed by offline analysis of the data by both the GstLAL and PyCBC pipelines~\citep{2020PhRvD.102b2004D, 2021PhRvD.104d2004C, 2022PhRvD.106j2006D, 2024PhRvD.110d3036K}. It was reported in \GRACEDB as S230814ah~\citep{GCNNOTICE}.

The inferred intrinsic astrophysical parameters are similar to those observed for a number of other events~\citep{LIGOScientific:2025slb}, including the first-ever \ac{GW} detection GW150914~\citep{2016PhRvL.116f1102A}.
We find good consistency between results obtained with different waveform models, as expected for this part of parameter space which is well covered by numerical and theoretical models.
The high \ac{SNR} of the signal enables us to perform a range of consistency tests between the data and theoretical expectations.
We performed many of the standard suite of tests described in \citet{GWTC4:TGR-I,GWTC4:TGR-II,GWTC4:TGR-III}, finding that while most are consistent with the expectation of \ac{GR} as described by the waveform models,
there is an apparent deviation in the post-merger portion of the signal that we investigate further here.
The dominant mode of the ringdown signal appears to decay faster than the expectation from the inferred remnant mass and spin.
Our follow-up investigations suggest that these deviations are likely due to either systematic errors in waveform models or detector noise effects.

This paper is structured as follows. 
In Section \ref{sec:basic_properties}, we present the detection of GW230814 and the measurement of its source properties. 
In Section \ref{sec:tgr_results} we present results of tests of \ac{GR} that have been performed on the event, finding that some
find an apparent deviation from \ac{GR} in the post-merger portion of the signal.
In Section \ref{sec:investigation} we detail further investigations to understand these results, focusing on data quality and waveform systematics.
Our conclusions are discussed in Section \ref{sec:conclusion}.

\ac{GW} observatory data, posterior samples and search information about GW230814 is available from The Gravitational Wave Open Science Centre~\citep{GWTC-4.0:Search,GWTC-4.0:PE}, and additional data used in the preparation of the paper is available at~\citet{zenodo}.

\section{Detection and Parameter estimates}\label{sec:basic_properties}

\subsection{Detector status and detection}

At the time of the signal, the \ac{LLO} was observing at a \acl{BNS} inspiral range~\citep{1993PhRvD..47.2198F} of approximately 155\,Mpc, consistent with its typical sensitivity during the observing run~\citep{2025PhRvD.111f2002C}.
The LIGO Hanford~\citep{2015CQGra..32g4001L}, Virgo~\citep{2015CQGra..32b4001A}, and KAGRA~\citep{2023PhRvX..13d1039A} detectors were not locked at the time of the event. 
Following standard \ac{GW} detection-vetting procedures~\citep{2025CQGra..42h5016S}, no data-quality issues were reported at the time of the event in the \ac{LLO} data. No sources of transient noise were found to be correlated with the gravitational-wave strain at the time of GW230814, based on data from auxiliary sensors.
Further details about the state of the detectors during O4a are provided in \citet{LIGOScientific:2025hdt}.

The signal was identified by the low-latency GstLAL pipeline~\citep{2017PhRvD..95d2001M, 2019arXiv190108580S, 2020PhRvD.101b2003H, 2021SoftX..1400680C, sakon2023template, ewing2023performance, 2023PhRvD.108d3004T} as a statistically significant deviation from the noise background, with a matched-filter SNR of \SNRgstlalonlinerounded\ and an \ac{IFAR} exceeding $10^5$ years.
The PyCBC Live~\citep{2016CQGra..33u5004U, 2017ApJ...849..118N, 2021ApJ...923..254D} 
and MBTA~\citep{2016CQGra..33q5012A, 2021CQGra..38i5004A, 2025CQGra..42j5009A} low latency pipelines are capable of detecting signals in single detector data, but did not identify the event, 
as they are configured to report only on single-detector signals with potential electromagnetic counterparts, typically those with template durations longer than 7\,s.
The event was not detected by search algorithms like coherent WaveBurst~\citep{2016PhRvD..93d2004K} and SPIIR~\citep{2022PhRvD.105b4023C}, which rely on coherence of signals in multiple detectors. 

\begin{table}[tb]
    \centering
    \begin{tabular}{l c c}
\hline \hline & GstLAL & PyCBC \\
\hline Offline \ac{SNR} & 42.3 & 43.0 \\
Online inverse FAR (yr) & $>10^{5}$ & - \\
Offline inverse FAR (yr) & $>10^{5}$ & $>1000$ \\
\hline \hline
\end{tabular}
\caption{\label{table:search} Significance and matched-filter SNR of GW230814 in two different search pipelines discussed in the text.}
\end{table}

In general, the \acp{IFAR} assigned to single-detector candidates carry greater uncertainty than those associated with multi-detector events because the background distribution is derived solely from triggers in one detector and cannot be combined with backgrounds from other instruments~\citep{2019arXiv190108580S,2020ApJ...897..169N,2022CQGra..39u5012C,2025CQGra..42j5009A}. The uncertainty is most important for marginally significant triggers near the tails of the background distribution, where statistical support is weak and constraints inherently poor. In contrast, triggers that fall within the bulk of the background and are consistent with noise, or those that are clearly separated from the noise distribution and consistent with a signal (like this event), can still be reliably identified~\citep{2020ApJ...892L...3A}.

Subsequent offline analysis using refined data-quality information and improved background estimation reaffirmed GstLAL's original significance estimate, also resulting in an \ac{IFAR} greater than $10^5$ years.
Furthermore, as summarised in Table~\ref{table:search}, the offline PyCBC search~\citep{2020PhRvD.102b2004D, 2021PhRvD.104d2004C, 2022PhRvD.106j2006D, 2024PhRvD.110d3036K}
also identified the event, assigning it an \ac{IFAR} exceeding 1000 years. The reported statistical significances reflect the maximum \ac{IFAR} permitted by each pipeline for single-detector events, as defined by their configuration choices. We therefore conclude with high confidence that this event is not of terrestrial origin.

\subsection{Source properties}
\begin{table}
	\begin{tabular}{l c} 
 \hline\hline
Parameter & Value \\ \hline
$m_1$ ($M_\odot$) & \massONEsourcecombined \\
$m_2$ ($M_\odot$) & \massTWOsourcecombined \\
Chirp Mass $\mathcal{M}$ ($M_\odot$) & \chirpmasssourcecombined \\
Total Mass $M$ ($M_\odot$) & \totalmasssourcecombined \\
Mass ratio $q = m_2 / m_1$ & \massratiocombined \\
$\chi_\mathrm{eff}$ & \chieffcombined \\
Final mass (detector frame) $M_\mathrm{f}^\mathrm{det}$ ($M_\odot$) & \finalmasscombined \\
Final spin $\chi_\mathrm{f}$ & \finalspincombined \\
$\cos(\theta_{JN})$ & \costhetajncombined \\
$\DL$ (Mpc) & \luminositydistancecombined \\
Redshift $z$ & \redshiftcombined \\
Optimal SNR & \networkoptimalsnrcombined \\
Matched filter SNR & \networkmatchedfiltersnrcombined \\
\hline\hline \end{tabular}

	\caption{\label{tab:pe} Selected parameters of the source averaged over the different waveform models used. 
    The value quoted is the median and the uncertainty indicates the 90\% credible interval.}
\end{table}

\Ac{PE} was performed using the \BILBY package~\citep{2019ApJS..241...27A, 2020MNRAS.499.3295R}, following an analysis setup consistent with \ac{gwtc4}~\citep{LIGOScientific:2025zdk} and four waveform models: \ac{seobnr}~\citep{2023PhRvD.108l4035P, Khalil:2023kep, vandeMeent:2023ols, 2023PhRvD.108l4037R}, \ac{xphm}~\citep{2021PhRvD.103j4056P, 2025PhRvD.111j4019C}, \ac{xo4a}~\citep{2021PhRvD.104l4027H, 2024PhRvD.109f3012T}, and \acf{nrsur}~\citep{2019PhRvR...1c3015V}.

The analysis used \reviewed{8 s of data 
around the trigger time, considering the frequency range
between $24$ Hz and $448$ Hz. This choice ensured that $\ell=|m|=3$ subdominant
modes were fully included in the analysis. }
The analysis marginalised over a detector calibration model that included frequency-dependent uncertainties in amplitude and phase. The calibration curves are consistent with zero offset in amplitude and phase with an uncertainty generally less than 3\% in amplitude and $2^\circ$ in phase across the bandwidth considered. The data specifying the calibration model is contained in the parameter estimation data release~\citep{GWTC-4.0:PE}.

The inferred signal, marginalized over the uncertainty in the source parameters,
is shown in Figure~\ref{fig:data}, which indicates the 90\% credible region for the \ac{nrsur}
waveform. We also show a wavelet reconstruction of the signal with the BayesWave algorithm~\citep{2015CQGra..32m5012C, 2017ApJ...839...15B, 2021PhRvD.103d4006C, bayeswave}, which we find is consistent with that from the \ac{nrsur} model. However, we observe that the BayesWave reconstruction has a quicker amplitude decay than the \ac{nrsur} reconstruction.

Table~\ref{tab:pe} shows the selected source parameters.
Unless otherwise noted, we report the parameter estimates averaged equally across waveform models, and quote the median and 90\% credible interval of the posterior probability distribution. Samples for individual waveform models are available in the \ac{gwtc4} data release~\citep{GWTC-4.0:PE}.

\begin{figure}
\includegraphics[width=\columnwidth]{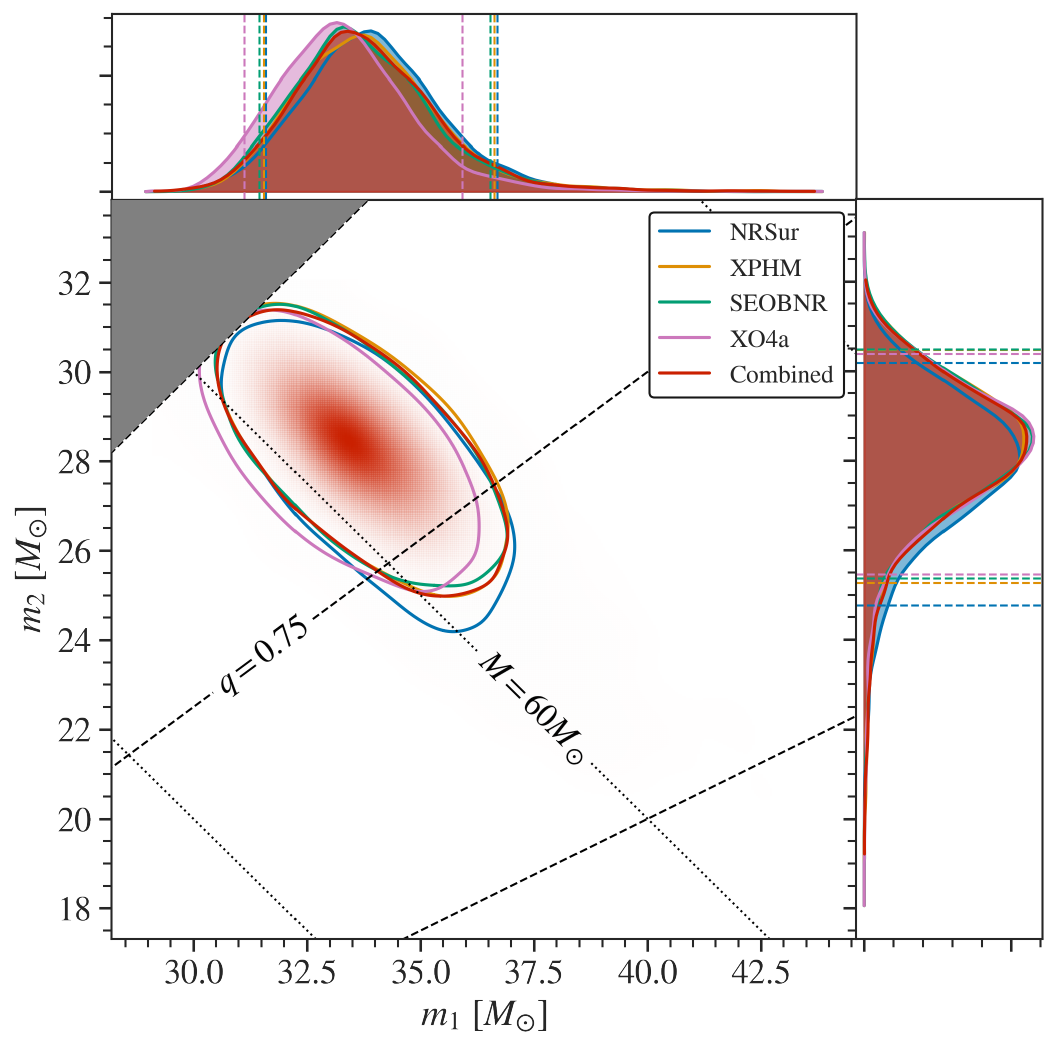}
\begin{caption}{\label{fig:masses}
The joint probability distributions for the two component masses, as measured in
the source frame using the four indicated waveform models. The shading density indicates
the combined posterior, averaging the posterior equally over the four waveform models.
Vertical dashed lines in the one-dimensional marginals indicate the $90\%$ credible intervals.
The individual component masses of the source are well-measured due to the high
\ac{SNR}, with values given in Table~\ref{tab:pe}.
}
\end{caption}
\end{figure}

In the source's rest frame, the system has a chirp mass $\Mc = \chirpmasssourcecombined~\Msun$, total mass $\Mtot = \totalmasssourcecombined~\Msun$, and components ($m_1$, $m_2$) = ($\massONEsourcecombined$, $\massTWOsourcecombined$) $~\Msun$.
The joint posterior distribution of the component masses is shown in Figure~\ref{fig:masses}. We find good agreement in the mass
measurements between the waveform models. 
All waveform models prefer a mass ratio away from the equal mass case, with an overall measurement $q=m_2/m_1=\massratiocombined$.

The luminosity distance was estimated as $\DL=\luminositydistancecombined$\,Mpc, implying a redshift $z=\redshiftcombined$
using the cosmological prior described in~\citet{2016A&A...594A..13P} and \citet{LIGOScientific:2025zdk}.
The detector-frame masses are related to source-frame masses by a factor of $(1+z)$. The merger remnant's detector-frame mass,
${\Mfdet}=\finalmasscombined\,\Msun$, 
and its final spin $\chi_\mathrm{f}=\finalspincombined$, determine the frequencies and damping
times of the \acp{qnm} (See Section~\ref{sec:tgr_results} for details).

\begin{figure}
    \includegraphics[width=\columnwidth]{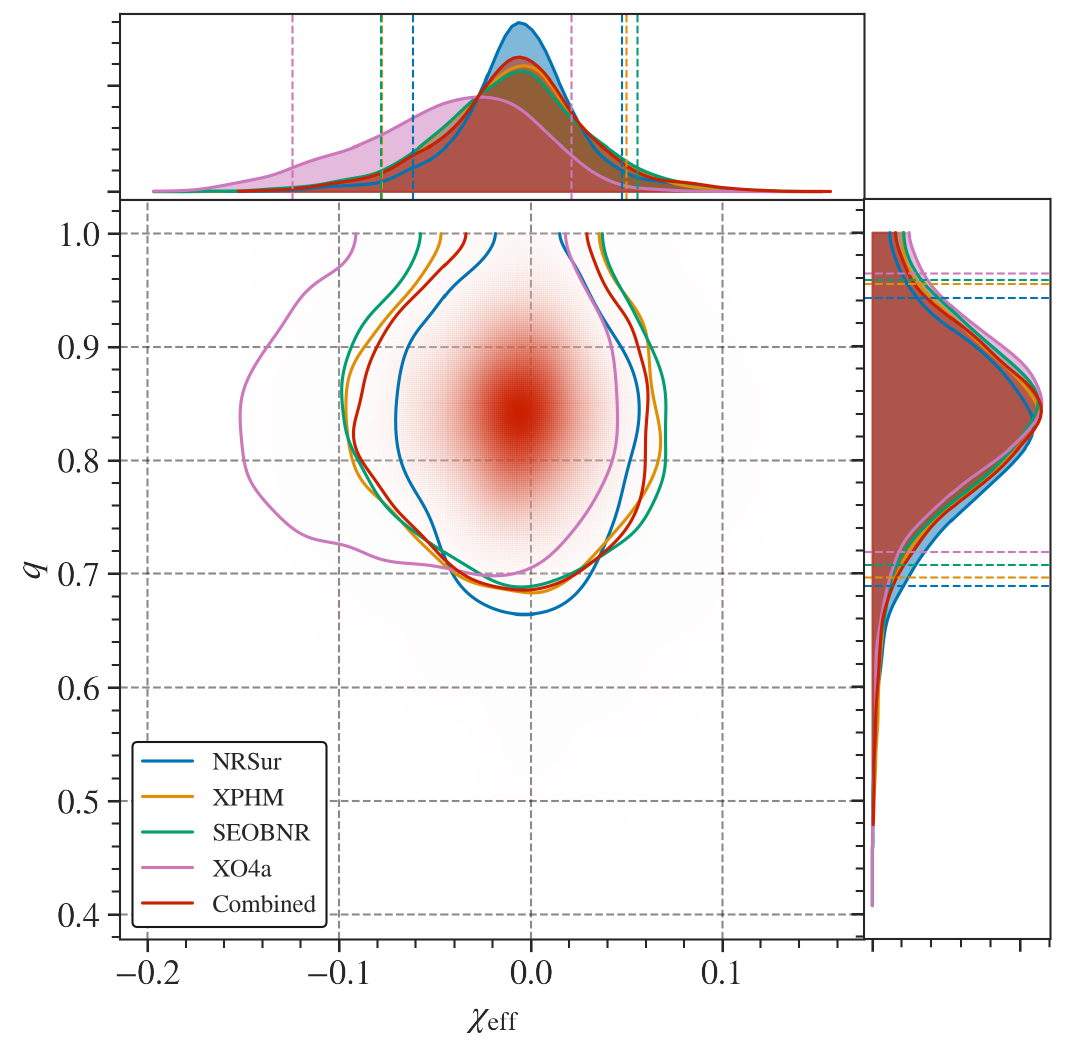}
    \caption{\label{fig:q_chi_eff} The joint probability distributions for the
    mass ratio $q$ and effective inspiral spin $\chieff$.  The mass ratio is well
    constrained away from equal masses, $q < 1$. The signal's effective inspiral spin is well measured to be
    $\chieff = \chieffcombined$ and is not strongly correlated with $q$.}
\end{figure}

\begin{figure*}
\includegraphics[width=0.24\textwidth]{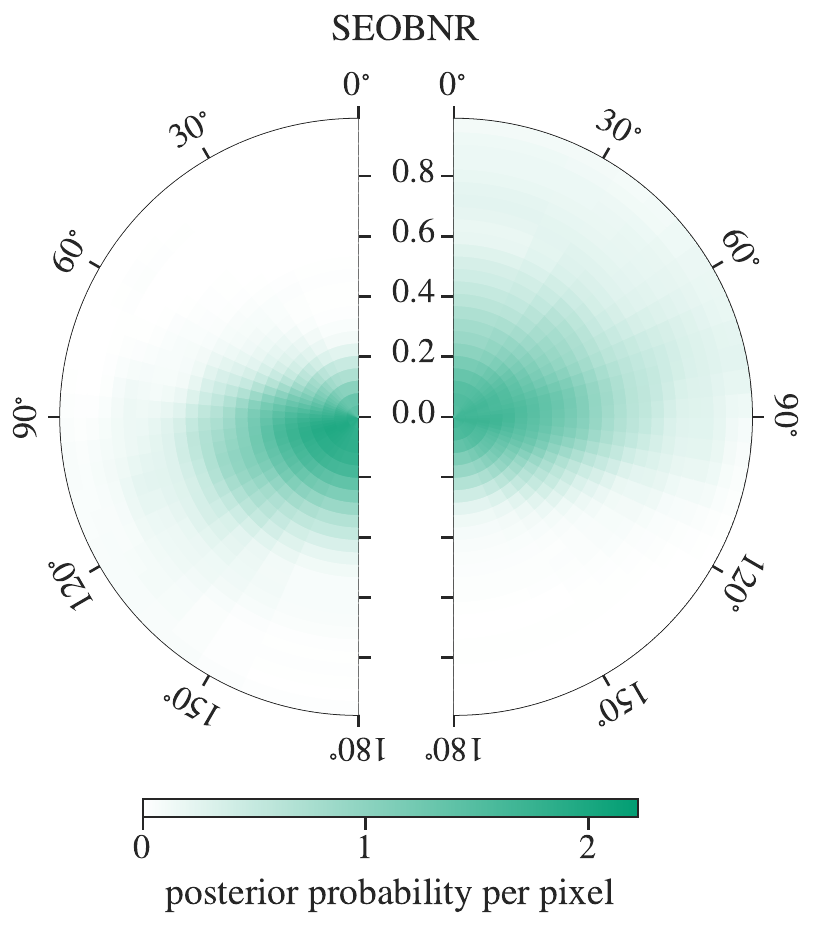}
\includegraphics[width=0.24\textwidth]{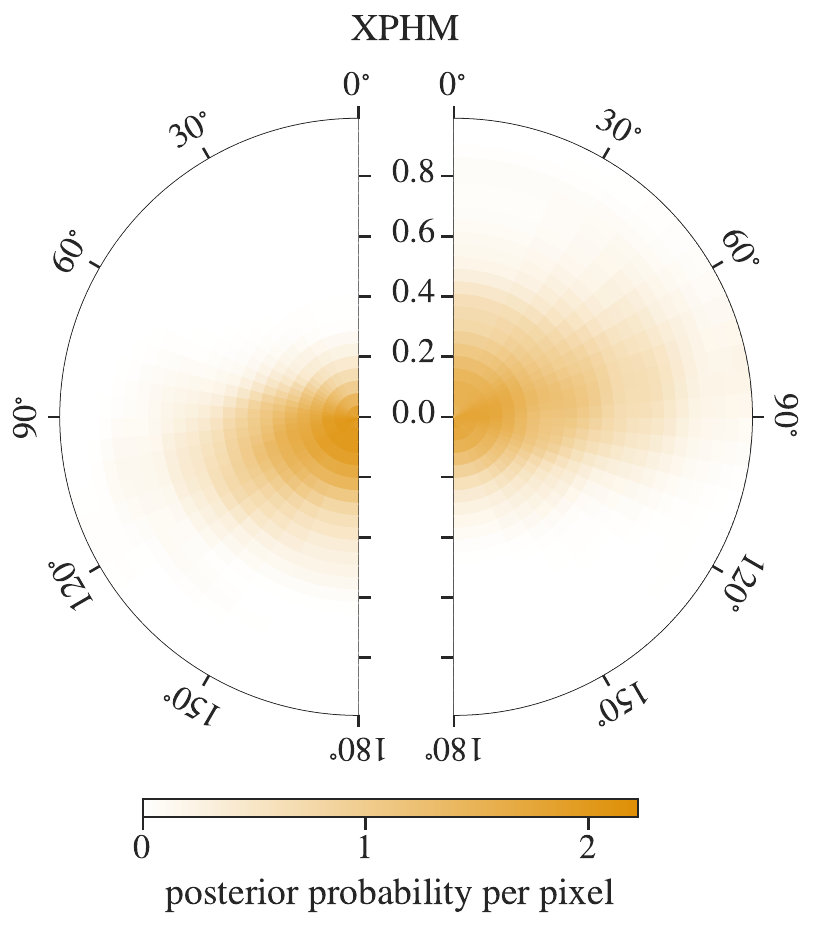}
\includegraphics[width=0.24\textwidth]{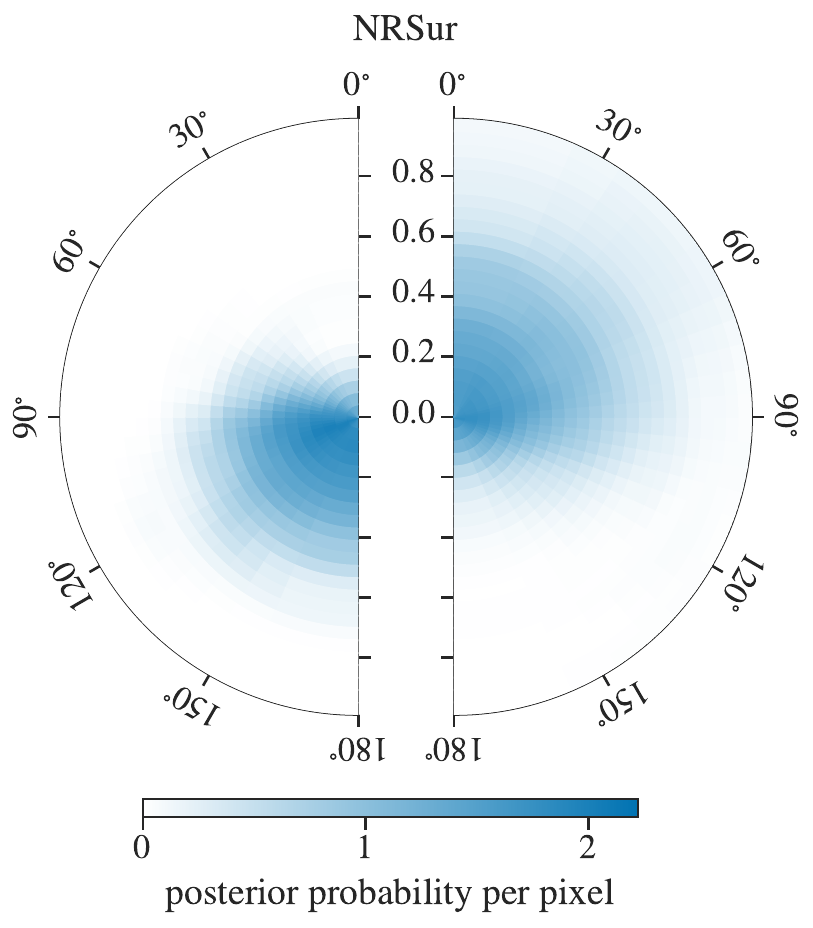}
\includegraphics[width=0.24\textwidth]{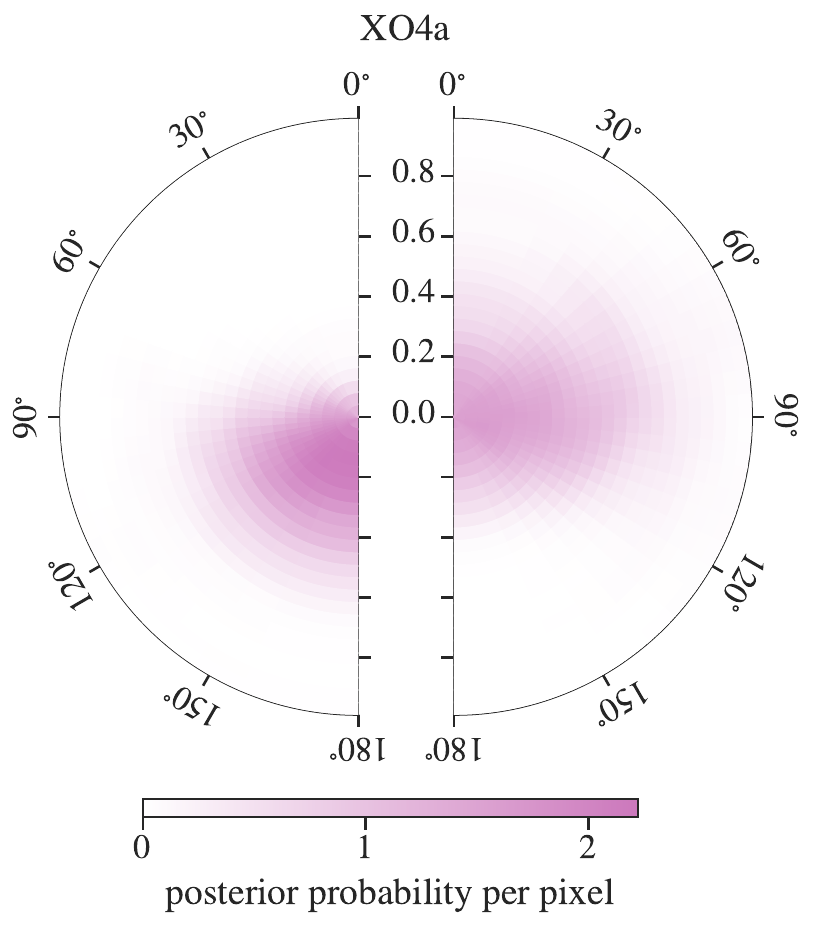}
\begin{caption}
{\label{fig:spin_disc} The joint posterior probability density estimates for the magnitude
and tilt angle of the spin vectors of the source, obtained using the four indicated waveform
models at a reference frequency of $f_\mathrm{ref}=20$\,Hz. The left and the right semicircles in each panel show the spins of the primary and secondary component objects, respectively. For all models, the spin vectors are constrained to have small magnitudes. 
}
\end{caption}
\end{figure*}

\begin{figure}
    \includegraphics[width=\columnwidth]{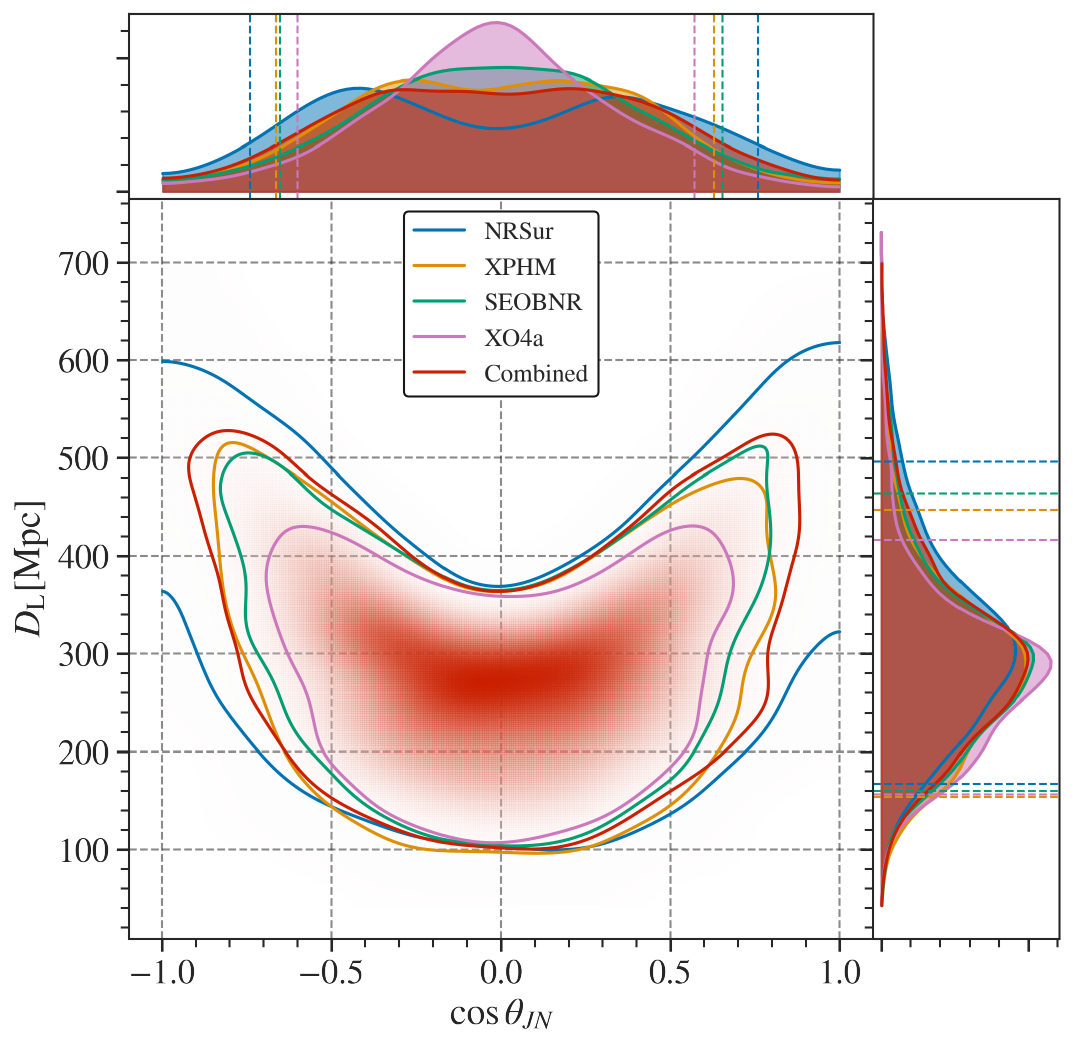}
    \caption{\label{fig:theta_jn} The joint posterior probability distribution for the cosine of the
    angle between the line of sight and the total angular momentum of the binary $\cos\theta_{JN}$
    at a reference frequency of \reviewed{20}\,Hz, and the luminosity distance to the source $\DL$.  These
    analyses rule out a face-on or face-off configuration at high credibility
    (the prior on $\cos \theta_{JN}$ is flat). At these signal strengths, despite the fact that a single detector 
    cannot constrain the polarization of the signal, the information on the viewing angle is provided by the sub-dominant 
    angular modes in the waveform. Credible intervals are given in Table~\ref{tab:pe}.}
\end{figure}

Given the high \SNR of the signal, we are able to constrain the spin vectors
of the compact objects well.
The best-measured spin parameter \chieff~\citep{2010PhRvD..82f4016S, 2009PhRvD..79h4032A}, defined as:
\begin{equation}
    \chi_{\mathrm{eff}} = \frac{ \vec{\chi}_1 \cdot \hat{L} + q \, \vec{\chi}_2 \cdot \hat{L} }{1 + q},
\end{equation}
is strongly constrained to 
\reviewed{$\chieff = \reviewed\chieffcombined$} and bears minimal correlation with the mass ratio as shown in Figure~\ref{fig:q_chi_eff}. Here, \(\vec{\chi}_1\) and \(\vec{\chi}_2\) are the dimensionless spin vectors of the primary and secondary compact objects respectively and \(\hat{L}\) is the unit orbital angular momentum vector, measured at a reference frequency of \reviewed{20}\,Hz.

Figure~\ref{fig:spin_disc} shows the magnitude of the spin vectors and their tilt angles relative to the orbital angular momentum vector of the system. 
The magnitudes of the spins are likely small, being constrained at 90\%
probability to be $\chi_1 < \reviewed{\aONEmagcombinedzerobound}$, $\chi_2 < \reviewed{\aTWOmagcombinedzerobound}$.
The spin components aligned with the orbital angular momentum direction are estimated to be $\vec{\chi}_1\cdot \hat{L}=\reviewed{\spinONEzcombined}$ and $\vec{\chi}_2\cdot \hat{L}=\reviewed{\spinTWOzcombined}$. The small inferred value of $\chieff$ implies that the aligned spin components are either small, or are negatively correlated such that the final $\chieff$ is close to zero.
The in-plane components of both spin vectors are found to be small and consistent with zero. Consequently, the effective precession parameter $\chip$~\citep{2015PhRvD..91b4043S}, which captures the dominant contribution from in-plane spins, is similarly constrained to be $\chip < \reviewed{\chipcombinedzerobound}$.

The angle $\theta_{JN}$ between the total angular momentum of the system and the line of sight is well-measured
(see figure~\ref{fig:theta_jn}), with results favoring edge-on viewing angles over face-on ones.
As the signal was observed by only a single detector, sensitive to only
one polarization, the entirety of this information must be coming from the relative multipole amplitudes and their differential 
emission patterns as functions of $\theta_{JN}$. When averaged over the different waveform models, 
the optimal \acp{SNR} $\rho_{\ell,m}$ of the most significant higher order angular $\ell,m$ multipoles of the inspiral are 
$\rho_{2,\pm 1}=\SNRTWOONE$, $\rho_{3,\pm 3}=\SNRTHREETHREE$, and $\rho_{4,\pm 4}=\SNRFOURFOUR$~\citep{2021PhRvD.103b4042M}.
\reviewed{The bound of the $\ell=|m|=4$ mode \ac{SNR} away from zero makes this the first \ac{CBC} with a
detection of an $\ell=4$ mode in the signal.}
This is further supported by a Bayes Factor of \(\log_{10} \mathrm{BF}\simeq 1 \) compared to a model with the $(4,\pm 4)$ multipoles excluded.
The \ac{SNR} rapidly and significantly decreases beyond these dominant few modes (\ac{SNR} $\leq 0.5$).

Despite the high \ac{SNR}, the measurement precision of source intrinsic parameters (masses and spins) is not extraordinarily high in the context of the overall catalog, as longer signals from lower mass sources tend to have more cycles in the sensitive band of the detector~\citep{1994PhRvD..49.2658C,1995PhRvD..52..848P}. Likewise, although GW230814 is relatively nearby, lower mass systems have been detected a similar distances with lower \ac{SNR}.

\subsection{Astrophysical context}

As the signal was only observed in a single detector, constraints on the sky location 
from timing information are absent. The remaining information from the prior that
favours higher distances make it unlikely that it originated from a direction that
corresponds to a lower response in \ac{LLO}.

\Ac{PE} indicates that the mass parameters of the GW230814 progenitor are consistent with those of the known \ac{BBH} population~\citep{2023PhRvX..13a1048A}.
\reviewed{
    Based on the known population, and the sensitivity of the detectors in O4, one would expect to observe a signal with matched-filter \ac{SNR} at least as great as that observed for GW230814 in approximately 1\% of systems.
The signal strength is therefore fully consistent with expectations for a population of the size found in GWTC-4.0~\citep{2025arXiv250818083T}}.

\section{Tests of General Relativity}\label{sec:tgr_results}

\begin{figure*}
    \centering
    \includegraphics[width=0.98\textwidth]{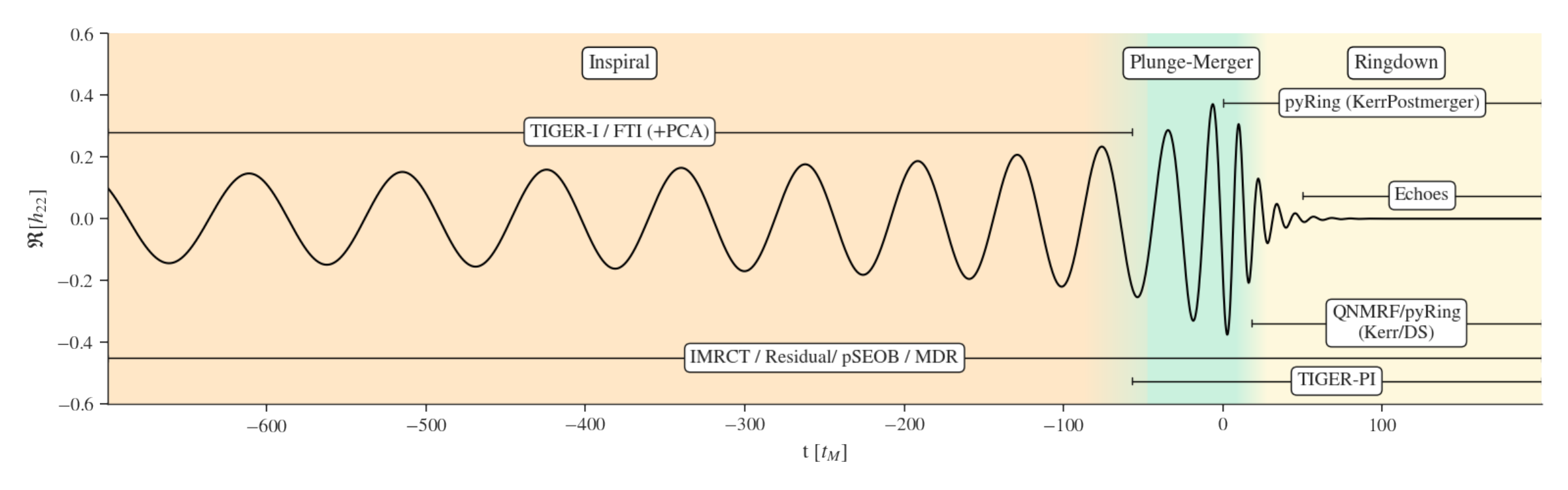}
    \caption{
        Illustrative $(2,2)$ mode waveform showing the domains of various tests of \ac{GR} performed on GW230814.
        The waveform is generated with typical parameters consistent with the posterior and is schematically
        divided into inspiral (orange), plunge--merger (light green), and ringdown (yellow). 
        Corresponding analysis frameworks include TIGER-I/FTI for inspiral, 
        TIGER-PI for plunge-merger-ringdown, pyRing (KerrPostmerger) for merger-ringdown, 
        QNMRF and pyRing (Kerr/DS) for ringdown, and IMRCT, residual, pSEOB, and MDR for the full signal. 
        Echo searches are performed in the late ringdown phase.}
    \label{fig:tgr-demo}
\end{figure*}

We present tests of \ac{GR} with GW230814, extending methods traditionally applied to coherent multi-detector observations—which suppress the impact of instrumental noise \citep{GWTC4:TGR-I,GWTC4:TGR-II,GWTC4:TGR-III}—to a single-detector signal.
The performed suite of tests probes the inspiral, ringdown, dispersion, and echoes, as qualitatively shown in Figure~\ref{fig:tgr-demo}. While most results are consistent with \ac{GR}, some outliers arise, particularly in the ringdown, as discussed below.

Of those that showed consistency with expectations, we briefly describe some null results here.
We performed a residuals test as described in~\citet{2025PhRvD.112h4080A}, that involves subtracting the maximum-likelihood
\ac{xphm} waveform from the data, analysing the residuals using BayesWave \citep{2015CQGra..32m5012C, 2021PhRvD.103d4006C, bayeswave},
and comparing it to a background of detector noise (the null hypothesis). This yields a p-value of $\RESIDUALP$,
showing consistency with noise, as under the null hypothesis, the p-values should be uniformly distributed between 0 and 1.
A Modified Dispersion Relation (MDR) test \citep{2012PhRvD..85b4041M} was conducted to probe 
the possibility of dispersion effects on the signal as it travels from the source to the detector. \reviewed{The results of this test 
are also consistent with \ac{GR}}. Finally, searches for both modelled \citep{Uchikata:2019frs,Lo:2018sep} 
and unmodelled \citep{Tsang:2018uie} late-time echoes were performed. Neither of these searches find any statistically significant evidence for echoes, \reviewed{returning $\log_{10}$ Bayes factors of $-0.2$ for the modelled search and $-4.1$ for the unmodelled search}.

\subsection{TIGER / FTI / PCA}\label{sec:fti-tiger}

Tests of the inspiral part of the waveform are performed by
both the Flexible Theory-Independent (FTI) \citep{Mehta:2022pcn} and 
Test Infrastructure for GEneral Relativity (TIGER) \citep{AgathosEtAl:2014, 2018PhRvD..97d4033M, Roy:2025gzv} pipelines.
These tests permit deviations from \ac{GR} in the non-spinning \ac{PN} coefficients, with each deviation parameter introduced and sampled independently, similar to the parameterized post-Einstein (ppE) framework \citep{2009PhRvD..80l2003Y,Cornish:2011ys}.
The FTI pipeline uses the \textsc{SEOBNRv5HM$\_$ROM} waveform model as its base \ac{GR} waveform, while TIGER relies on \ac{xphm}.
The deviations are applied to the non-spinning part of the \ac{PN} coefficients in \ac{GR} $\psi_i^\mathrm{GR, NS}$, as
\begin{equation}
    \psi_i \rightarrow (1+\dphi{i})\psi_i^\mathrm{GR, NS} + \psi_i^\mathrm{GR, S},
    \label{eq:frac_PN_dev}
\end{equation}
where $\dphi{i}$ are the deviation parameters and $(\psi_i^\mathrm{GR, NS}, \psi_i^\mathrm{GR, S})$ 
are the spin-independent and spin-dependent part of the \ac{PN} coefficient from \ac{GR}, respectively.
These corrections are applied in the inspiral part of the signal and then turned off when transitioning to the merger-ringdown.
FTI and TIGER differ mainly in their choice of cutoff frequency: FTI adopts a higher cutoff, allowing it to probe deeper into the late inspiral and thereby place tighter constraints on the high-order \ac{PN} coefficients.

The TIGER pipeline additionally modifies coefficients in the
intermediate and merger-ringdown parts of the waveform. 
Parameters ${\delta \hat{b}_1, \delta \hat{b}_2, \delta \hat{b}_3, \delta \hat{b}_4}$ capture deviations in the \ac{NR}-calibrated coefficients 
in the intermediate regime, and the parameters ${\delta \hat{c}_1, \delta \hat{c}_2, \delta \hat{c}_4, \delta \hat{c}_\ell}$ capture deviations 
in the merger-ringdown calibration coefficients~\citep{Roy:2025gzv}.

\begin{figure}
    \centering
    \includegraphics[width=0.98\linewidth]{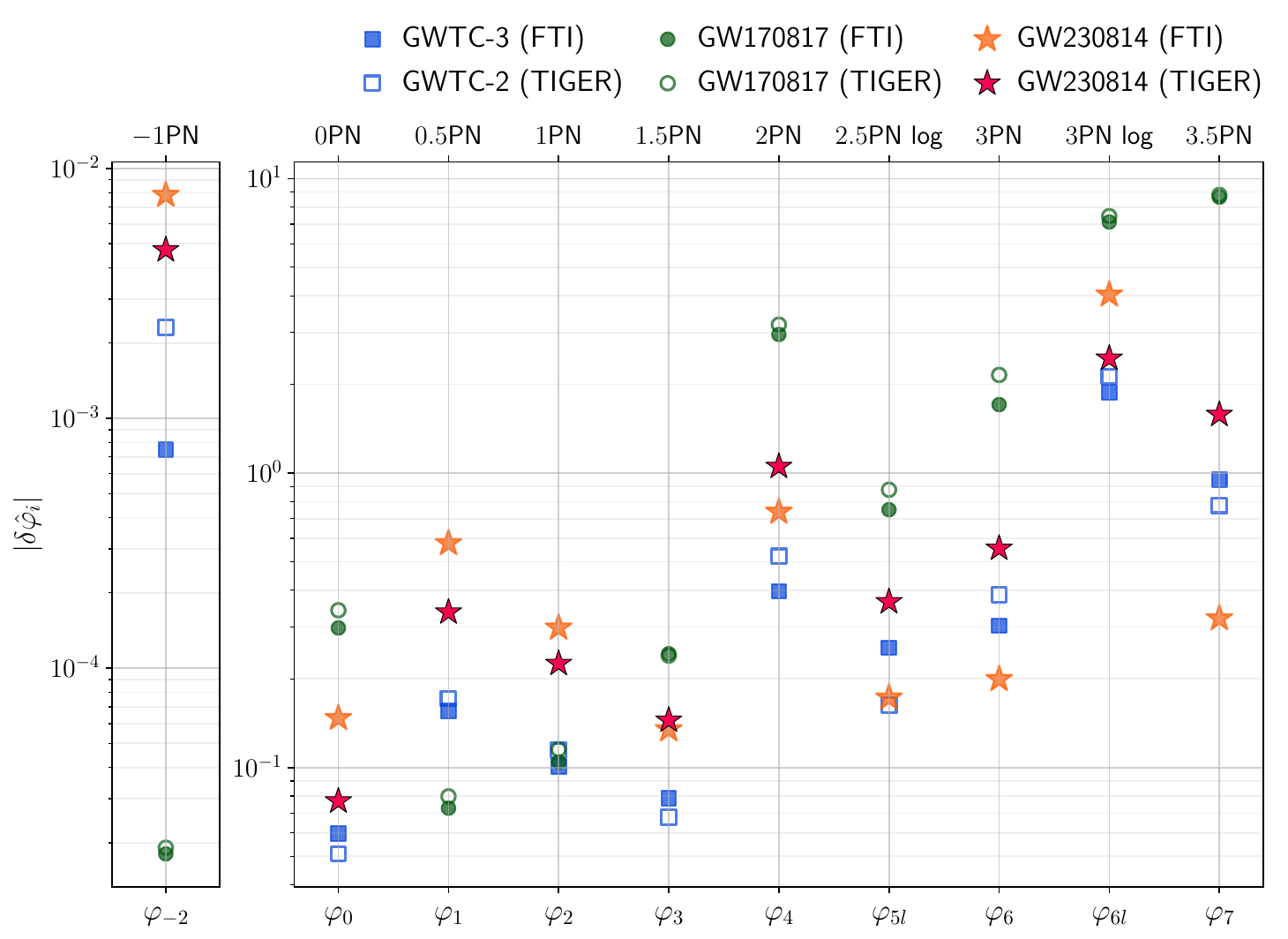}
    \caption{Bounds from the FTI and TIGER tests on the PN coefficients in the inspiral part of
    the signal. For higher-order PN terms, the bounds improve on similar previously presented
    constraints from GW170817~\citep{LIGOScientific:2018dkp}, and in some cases, are tighter than bounds from the
    entire GWTC-2 and GWTC-3 populations \citep{2021PhRvD.103l2002A, 2025PhRvD.112h4080A}.
    The FTI analysis employs uniform priors in the range $[-1, 1]$ for $\delta{\hat{\chi}}_{-2}$ and $[-20, 20]$ for all other parameters.
    The TIGER analysis employs uniform priors in the ranges $[-0.1, 0.1]$ for $\delta{\hat{\chi}}_{-2}$; $[-5.0, 5.0]$ for $\delta{\hat{\chi}}_{0}$; $[-2.0, 2.0]$ for $\delta{\hat{\chi}}_{1}$, $\delta{\hat{\chi}}_{2}$ and $\delta{\hat{\chi}}_{3}$; $[-20.0, 20.0]$ for $\delta{\hat{\chi}}_{4}$, $\delta{\hat{\chi}}_{5l}$ and $\delta{\hat{\chi}}_{6}$; and $[-30.0, 30.0]$ for $\delta{\hat{\chi}}_{6l}$ and $\delta{\hat{\chi}}_{7}$.}
    \label{fig:fti_comp}
\end{figure}

\begin{figure}
    \centering
    \includegraphics[width=0.98\linewidth]{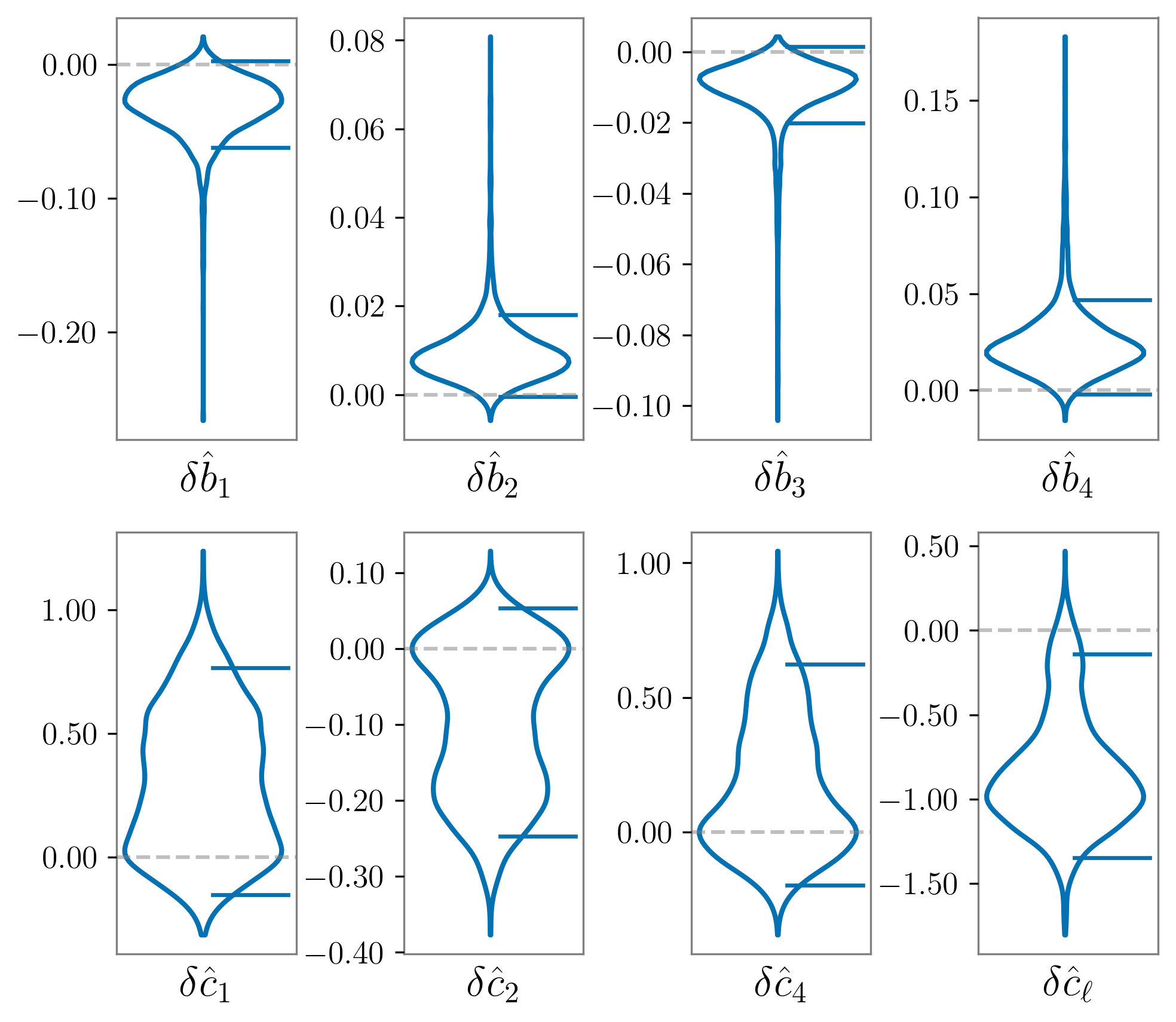}
    \caption{Results from the TIGER test on the intermediate and merger-ringdown part of the signal using the waveform \ac{xphm}.
    The red dashed lines indicate the GR prediction at $0$. 
    For the intermediate coefficients $\delta \hat{b}_i$, the GR prediction value of zero lies within the 90\% credible intervals, but for the merger-ringdown coefficient, $\delta \hat{c}_l$, the \ac{GR} prediction lies outside the 90\% credible interval.
    For each of these intermediate and merger-ringdown coefficients, the TIGER test employs uniform priors in $[-20,20]$.}
    \label{fig:tiger_result}
\end{figure}

Figures~\ref{fig:fti_comp} and \ref{fig:tiger_result} summarize the results. The inspiral parameters are consistent with \ac{GR}, with Figure~\ref{fig:fti_comp} showing 90\% upper bounds on the magnitude of deviations. Differences between FTI and TIGER bounds arise primarily from their cutoff frequency choices and reflect the systematic uncertainties associated with modeling. Because the FTI test for GW230814 uses a higher tapering frequency than in earlier implementations, these bounds are not directly comparable with those reported in GWTC-3. In contrast, the bounds from GW170817~\citep{LIGOScientific:2018dkp} remain comparable, as the old and new tapering frequencies lie outside the detectors’ sensitive band.

During the inspiral, both FTI and TIGER are consistent with \ac{GR} (deviation parameters consistent with zero), whereas the TIGER's merger--ringdown parameter $\delta \hat{c}_l$ has a posterior shifted from the \ac{GR} prediction. The \ac{GR} quantile for this $\delta \hat{c}_l$ parameter is \reviewed{$2.4\%$}. Despite this shift, we find no statistically significant evidence for a deviation from \ac{GR}, with \reviewed{$\log_{10}$ Bayes factors, $\log_{10} \mathcal{B}$ of $-0.6$} in favor of \ac{GR}.
We note that these values depend on prior choices; in the TIGER pipeline, each deviation parameter for the intermediate and merger-ringdown phase is assigned a uniform prior in $[-20,20]$, chosen conservatively to match the ranges used for quieter signals with broader posteriors.

The FTI and TIGER pipelines vary each deviation parameter separately.
To account for possible correlated deviations across the inspiral parameters, starting at 1.5PN order 
and including up to 3.5PN order, we diagonalize the covariance matrix 
and identify the leading and sub-leading eigenvectors (linear combinations of the original 
\ac{PN} deformation parameters, referred to as Principal Component Analysis (PCA) parameters) that have the smallest 
eigenvalues~\citep{Pai:2012mv,Saleem:2021nsb,Shoom:2021mdj,Mahapatra:2025cwk}.
In this parametrization, results for the PCA components from both the TIGER and the FTI frameworks are consistent with \ac{GR}, with zero deviation lying within the 90\% credible intervals of the posterior.

\subsection{Inspiral-Merger-Ringdown Consistency}\label{sec:imr-consistency}

The \ac{IMRCT} \citep{Hughes:2004vw,2016PhRvD..94b1101G,2018CQGra..35a4002G} compares the consistency of the final mass and spin inferred from the data below and above a frequency cutoff that approximately separates the inspiral and merger--ringdown parts of the signal in the frequency domain.
The frequency cutoff of $\IMRCTCUT$\,Hz is set to be the innermost circular orbit frequency of the remnant Kerr black hole whose properties are inferred with the entire signal using the \ac{xphm} waveform that is used for this test.
The \acp{SNR} for the signal below and above the cutoff are $\IMRCTLOWSNR$ and $\IMRCTHIGHSNR$.

\begin{figure}
    \centering
    \includegraphics[width=0.98\linewidth]{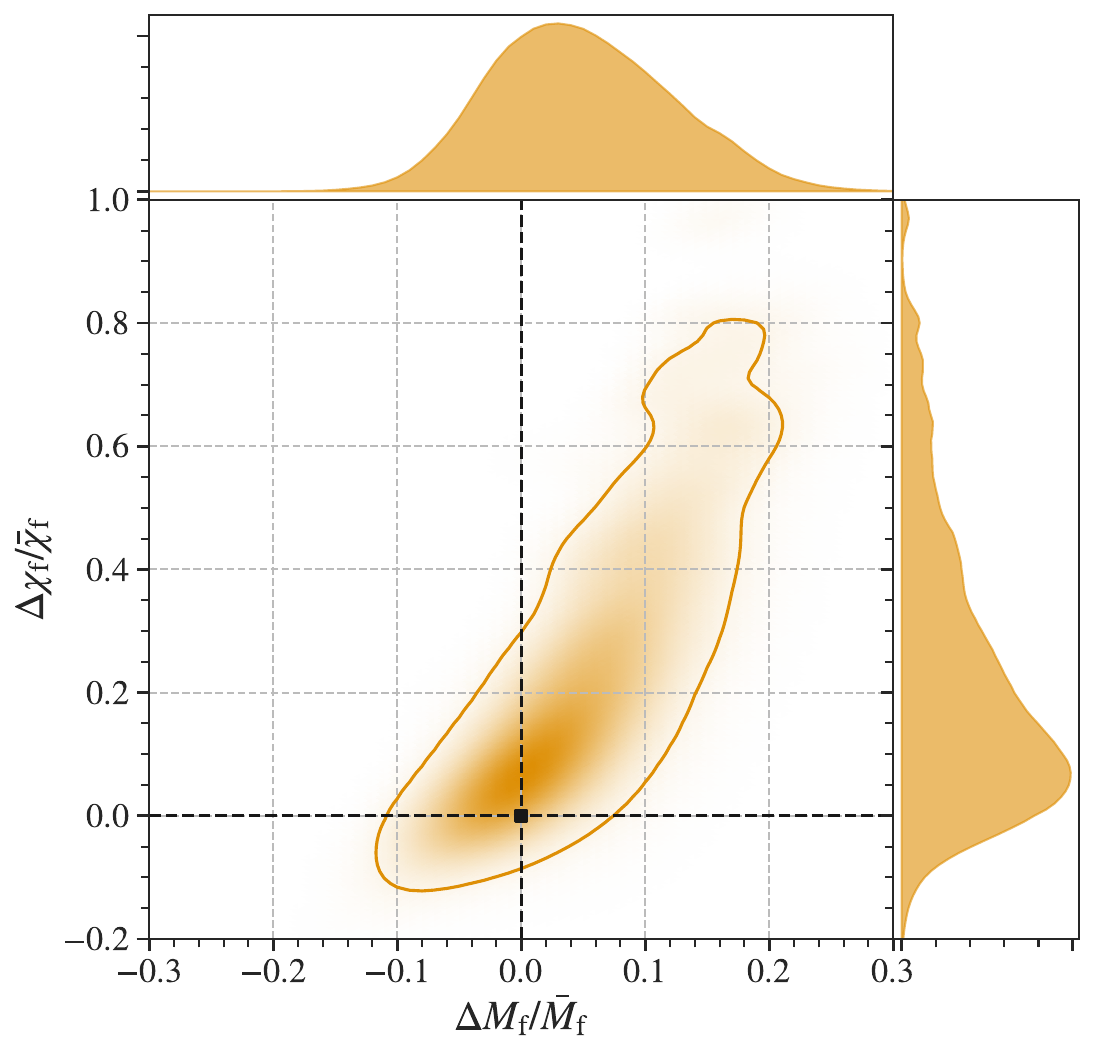}
    \caption{Results from the \ac{IMRCT} using the waveform \ac{xphm}. The black square denotes the GR prediction at $\bar{M}_{\rm f} = \bar{\chi}_{\rm f} = 0$. The \ac{GR} parameters of the 
    lower-frequency and higher-frequency parts of the signal are seen to be consistent with this test.}
    \label{fig:imrct_result}
\end{figure}

Figure~\ref{fig:imrct_result} shows the results of the test in terms of the  parameters
\begin{eqnarray}
\frac{\Delta M_{\rm f}}{\bar{M}_{\rm f}} &= 2 \frac{M_{\rm f}^{\rm low}-M_{\rm f}^{\rm high}}{M_{\rm f}^{\rm low}+M_{\rm f}^{\rm high}} \,, \quad \frac{\Delta \chi_{\rm f}}{\bar{\chi}_{\rm f}} = 2 \frac{\chi_{\rm f}^{\rm low} - \chi_{\rm f}^{\rm high}}{\chi_{\rm f}^{\rm low} + \chi_{\rm f}^{\rm high}},
\end{eqnarray}
where $\bar{M}_{\rm f}$ and $\bar{\chi}_{\rm f}$ denote the averages of the final mass and spin estimated from the two parts of the signal analyzed separately.
Following~\citet{2025PhRvD.112h4080A}, we also reweigh the \ac{IMRCT} posteriors in the final mass and spin deviation parameters to account for the non-uniform priors 
induced by the mapping from uniformly sampled component masses.

The results appear consistent with the \ac{GR} prediction of $(0,0)$, which is contained within the \reviewed{47\%} contour.
This indicates \ac{IMRCT} is not sensitive to the anomaly found by TIGER in the merger-ringdown.
The credible region is comparable in size to that for the entire population in \ac{gwtc3}~\citep{2025PhRvD.112h4080A}.

\subsection{pSEOBNR}\label{sec:pseobnr}

The \ac{pseobnr} analysis~\citep{Brito:2018rfr, Ghosh:2021mrv, Maggio:2022hre, Toubiana:2023cwr, Pompili:2025cdc} introduces
fractional deviations $\delta \hat{f}_{\ell m 0}$ and $\delta \hat{\tau}_{\ell m 0}$ in the
frequency and damping time of the dominant \acp{qnm} in the ringdown portion of the \ac{seobnr} waveform as:
\begin{eqnarray} 
    f_{\ell m 0} = f_{\ell m 0}^{\text{GR}} (1 + \delta \hat{f}_{\ell m 0}), \, \,
    \tau _{\ell m 0} = \tau_{\ell m 0}^{\text{GR}} (1 + \delta \hat{\tau}_{\ell m 0}),
\end{eqnarray}
where the \ac{GR} values depend on the remnant's mass and spin~\citep{Stein:2019mop}, which are inferred from the progenitor binary's
properties via \ac{NR} fits~\citep{2017PhRvD..95f4024J,2016ApJ...825L..19H}. The indices $(\ell m n)$ label the \ac{qnm} angular harmonics $(\ell m)$ and overtones ($n=0$ being fundamental).
This approach employs the full signal length and \ac{SNR} as \ac{pseobnr} modifies these parameters within a full \ac{IMR} model.
The present analysis targets deviations in the $(2,2,0)$ \ac{qnm} by jointly sampling over $(\delta \hat{f}_{220}, \delta \hat{\tau}_{220})$,
along with the standard \ac{BBH} parameters. For the \ac{qnm} deviation parameters, we use uniform priors in the ranges $\delta \hat{f}_{220} = [-0.8,2.0]$ and $\delta \hat{\tau}_{220} = [-0.8,2.0]$.
The other modes included in the model, but not modified, are the $(2,1,0), (3,3,0), (3,2,0), (4,4,0), (4,3,0)$ and $(5,5,0)$~\citep{2023PhRvD.108l4035P}.

\begin{figure}
    \centering
    \includegraphics[width=0.98\linewidth]{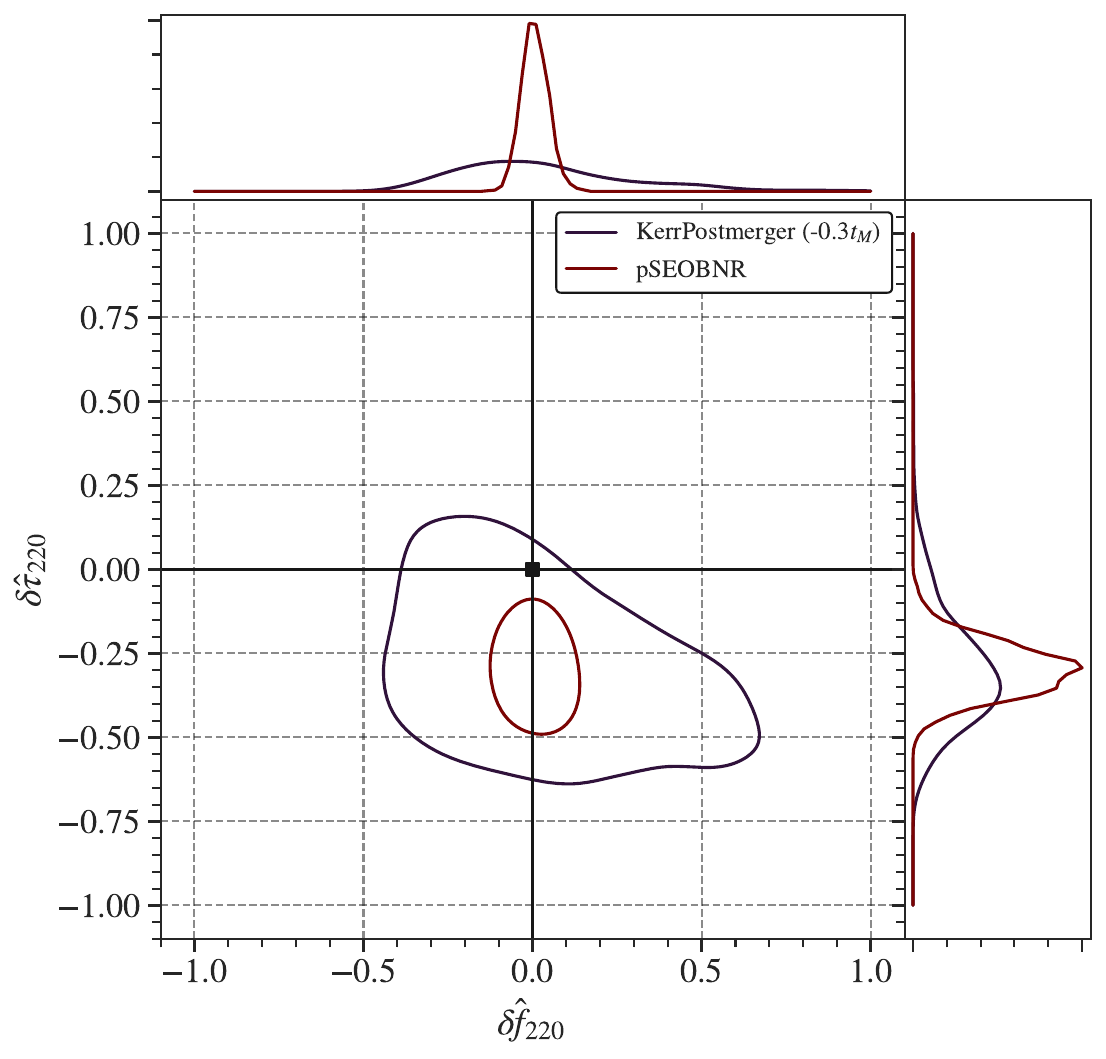}
    \caption{The 90\% credible regions for the posterior probability distribution of the fractional deviations in the frequency
and damping time of the $(2,2,0)$ \ac{qnm}, $(\delta \hat{f}_{220}, \delta \hat{\tau}_{220})$ for GW230814. 
The contours and corresponding marginalized posteriors are shown for the \ac{pseobnr} (dashed--dotted maroon, see Section~\ref{sec:pseobnr}) and pyRing KerrPostmerger (solid black, see Section~\ref{sec:pyRing})  analyses. The black square denotes the \ac{GR} prediction of $\delta \hat{f}_{220} = \delta \hat{\tau}_{220} = 0$.}
    \label{fig:pSEOB_result}
\end{figure}

The results of the analysis of GW230814 are summarized in Figure~\ref{fig:pSEOB_result}, which shows the 90\% credible levels of the posterior probability distribution of the fractional deviations in the frequency and damping time of the $(2,2,0)$ \ac{qnm}, and their corresponding one-dimensional marginalized posterior distributions.
The data support a damping time smaller than expected from \ac{GR}. The bounds from the \ac{pseobnr} analysis are \reviewed{$\delta \hat{\tau}_{220} = -0.3^{+0.15}_{-0.12}$},  
with a \ac{GR} quantile of \reviewed{$1-\rm{CI}(\delta \hat{{\it f}}_{220}=\delta \hat{\tau}_{220} = 0)=0.0092$} where $\rm{CI}$ denotes the credible interval.
Bayesian model selection between the \ac{pseobnr} and \ac{seobnr} models results in a $\log_{10} \mathcal{B}$ of \reviewed{$-0.43$} between the two models, indicating that
the data favours \ac{seobnr}. However, as mentioned before, 
Bayes factors are sensitive to the choice of prior ranges for the deviation parameters.
The credible-level result indicates potential deviations from the \ac{GR} prediction.
The result is in tension with the population constraint from \ac{gwtc3}, which is $\delta \hat{\tau}_{220} = 0.13^{+0.21}_{-0.22}$~\citep{2025PhRvD.112h4080A}.

The measured frequency and damping time from the \ac{pseobnr} result are shown in Figure~\ref{fig:f-tau}, in comparison
with results from other analyses described below. The \ac{IMRCT} and residuals test mentioned earlier, being broadband in nature, may not
be as sensitive as \ac{pseobnr} to specific deviations in the waveform. The consistency between the \ac{pseobnr} and other results is
further strengthened by running the TIGER and \ac{IMRCT} pipelines on the maximum-likelihood waveform from the \ac{pseobnr} analysis.
These tests yield \reviewed{similar results} to the actual signal, with TIGER finding the deviation and \ac{IMRCT} not.

\subsection{pyRing}\label{sec:pyRing}

pyRing~\citep{Carullo:2019flw, pyRing} analyzes the post-merger phase of \ac{BBH} coalescences using a time-domain likelihood~\citep{Carullo:2019flw, Isi:2021iql, DelPozzo:2016kmd}. Unlike the \ac{pseobnr} pipeline, it excludes the inspiral portion of the signal. The analysis employs multiple signal models, each suited to different times after merger, with time conveniently expressed in units scaled by the detector-frame final mass, $t_M = G \Mfdet / c^3 \simeq 0.31\,\mathrm{ms}$. To search for possible deviations from \ac{GR}, the framework allows for parameterized shifts in the Kerr \ac{qnm} frequencies and damping times, determined by the final detector-frame mass and spin of the remnant \ac{BH}.

We adopt the median of the posterior of the invariant strain peak defined as:
\begin{equation}
    t_\mathrm{peak} = \max_t \sqrt{\sum_{\ell, m} \left| h_{\ell m}(t) \right|^2}
\end{equation}
from the
\ac{nrsur} analysis in \ac{LLO} as our fiducial reference GPS time $t_{\rm LLO} = \reviewed{1376089759.805280}{}$\,s, and report offsets from this time in mass units $t_M$.

The most agnostic \ac{ds} model represents 
the ringdown portion of the waveform as a sum of decaying sinusoids with free 
amplitudes and phases. Since this model allows for only \ac{qnm} with
time-independent frequencies and damping times, the analysis is started 10\,$t_M$ after $t_\mathrm{LLO}$~\citep{Kamaretsos:2011um}.
The Kerr model builds on this by incorporating the 
frequency structure associated with perturbations of Kerr \acp{BH}~\citep{Berti:2014fga}.
The KerrPostmerger model~\citep{Gennari:2023gmx},
includes information about the binary progenitor system, relying on time-dependent amplitudes calibrated to \ac{NR} simulations, which allow it to fit the entire post-peak data~\citep{Damour:2014yha, DelPozzo:2016kmd, Nagar:2019wds, 
Nagar:2020pcj}. The model includes the $(\ell, |m|)=$ $(2,2)$, $(2,1)$, $(3,3)$, $(4,4)$, and $(5,5)$ angular multipoles, and allows deviations in the frequency and damping time of the $(2,2)$ mode.
Among all signal models, 
KerrPostmerger is the most sensitive, as it effectively captures the non-linearities, overtone excitations, and transient features near the merger, allowing for the most detailed extraction of information from the post-merger signal, starting at the peak of the (2,2) invariant strain.

Results from the \ac{ds} model containing a single sinusoid component starting
at \reviewed{$10\,t_M$} after $t_{\rm LLO}$ yield a measurement of frequency $f =
\reviewed{246^{+24}_{-32} \, \mathrm{Hz}}{}$ and damping time $\tau = \reviewed{2.80^{+1.27}_{-0.82} \, \mathrm{ms}}{}$.
These results are illustrated in Figure~\ref{fig:f-tau}. Using the remnant properties
estimated from the \ac{nrsur} \ac{IMR} analysis, the predicted \ac{qnm} frequencies and
damping times can be computed for a subset of modes and compared to these
agnostic measurements. The frequency and damping time of the
\ac{ds} do not completely overlap with those of the $(\ell,m,n) = (2,2,0)$ 
Kerr fundamental mode of a remnant black hole with properties inferred from 
our \ac{IMR} analysis. Instead, the median frequency of the \ac{ds} lies between those of the $(2,2,0)$ and $(2,1,0)$ modes, as shown in Figure~\ref{fig:f-tau}.

Using the Kerr model, with the $(2,2,0)$ and $(2,1,0)$ modes included, we find
that the damping time is reduced relative to the single \ac{ds}, and is
closer to the results from the \ac{pseobnr} analysis. The frequencies inferred
for the individual modes are set by the common $\Mfdet$ and $\chi_f$ parameters in the model, and are consistent with the \ac{IMR} result, however the damping times show a lower than expected value.

Results for the KerrPostmerger model, which analyses data from times before the damped
sinusoid model, are also shown in Figure~\ref{fig:f-tau}. The frequency and damping time for
the KerrPostmerger model also do not completely overlap with those of the $(2,2,0)$ 
Kerr fundamental mode of a remnant black hole with properties inferred from 
our \ac{IMR} analysis. This time, the damping time lies largely below
that of the $(2,2,0)$ mode, consistent with the finding from the \ac{pseobnr} analysis, although the $90\%$ credible region is somewhat larger.
The discrepancy between the \ac{GR} prediction and the KerrPostmerger fit is quantified by the deviation parameters $\delta \hat\tau_{220}$ and $\delta \hat f_{220}$, with 90\% credible interval shown in Figure~\ref{fig:pSEOB_result}.
Bayesian model selection between the KerrPostmerger model with deviation parameters
allowed to vary in the range $\pm 100\%$ and the same model with $\delta \hat\tau_{220} = \delta \hat f_{220} =0$ yield a $\log_{10} \mathcal{B}$ of \reviewed{1.9} in favour of the GR model.

\subsection{QNM Rational Filter}\label{sec:QNMRF}
The \acp{qnmrf} pipeline~\citep{Ma:2022wpv,Ma:2023vvr,Ma:2023cwe, Lu:2025mwp} analyzes the post-merger signals from \ac{BBH} coalescences by applying a frequency-domain filter to remove signal components with specific complex-valued frequencies. This approach can eliminate one or more \acp{qnm} corresponding to a given remnant \ac{BH} mass $\Mfdet$ and spin $\chi_\mathrm{f}$ from the ringdown signal. The filtered waveform is then transformed back to the time domain and analyzed using the same time-domain likelihood formalism as described in Section~\ref{sec:pyRing}. Since the filters remove \acp{ds} or \acp{qnm} without requiring prior knowledge of their amplitudes or phases, the likelihood function remains two-dimensional regardless of the number of modes removed. The \acp{qnmrf} analysis adopts the same starting-time convention as that in Section~\ref{sec:pyRing} with the same $t_M$ and $t_{\rm LLO}$ values.

\begin{figure*}
    \begin{center}
        \includegraphics[width=2\columnwidth]{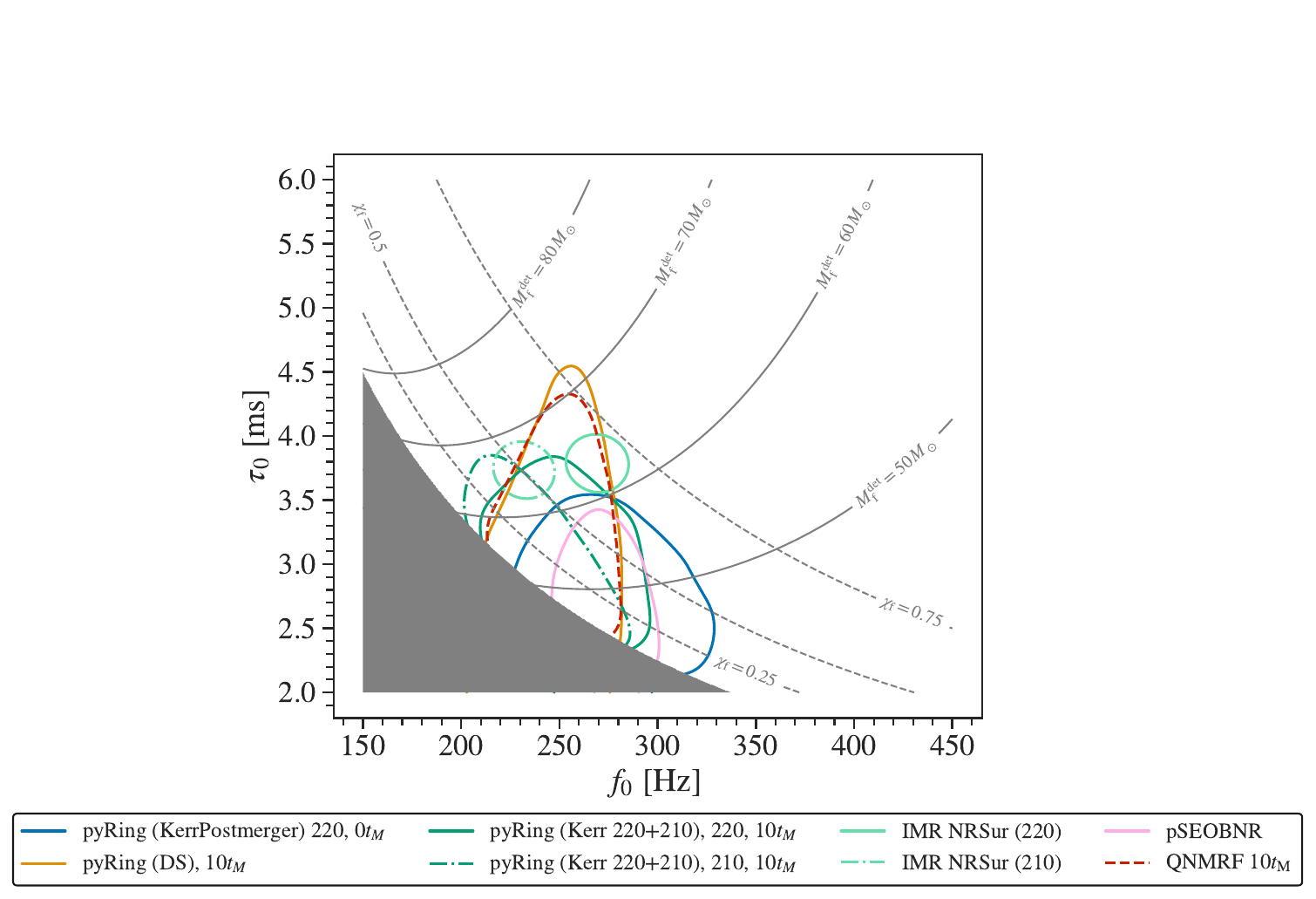}
    \end{center}
    \caption{\label{fig:f-tau} Inferred frequencies and damping times of \ac{ds}s
    or modes from models discussed in the text, see section
    \ref{sec:tgr_results}. The light gray lines show the frequency and
    damping time of the $(2,2,0)$ fundamental mode of a Kerr black hole at
    various masses and spins. The dark gray region corresponds to the 
    frequencies and damping times that are inconsistent with a Kerr $(2,2,0)$
    fundamental mode. The light green solid and dashed curves show the
    frequency and damping time of the Kerr $(2,2,0)$ and $(2,1,0)$
    fundamental modes for the remnant properties measured in our \ac{IMR}
    analysis with the \ac{nrsur} waveform model. We show results from
    an analysis using the \ac{pseobnr} (Section \ref{sec:pseobnr}) model in pink.
    The inferred frequency and damping time of the $(2,2,0)$ fundamental mode
    in this analysis (also shown in Figure~\ref{fig:pSEOB_result}) are
    inconsistent with the same mode using estimates of the remnant properties from
    our \ac{IMR} analysis. Analysis of the post-peak signal starting at 
    \reviewed{$t \sim 10\,t_M$} using the pyRing pipeline with a single \ac{ds}
    (Section \ref{sec:pyRing}) is shown in dark orange.
    The frequency from this analysis spans those inferred by the \ac{IMR}
    analysis for the $(2,2,0)$ and the $(2,1,0)$ modes, and the damping time is consistent with lower values.
    Analysis with the Kerr model, using both the (2,2,0) and the (2,1,0) modes is shown in dark green, 
    and shows that assuming the two modes are present in the signal finds a lower damping time, more consistent with the \ac{pseobnr} result.
    Analysis of the post-peak signal using the {KerrPostmerger} model 
    (Section \ref{sec:pyRing}) starting at \reviewed{$t = 0\,t_M$} 
    are shown in blue. This model also finds a frequency and damping time for the $(2,2,0)$ Kerr fundamental mode that are in tension with the \ac{IMR} results. Analysis of the post-peak signal starting at $t \sim 10\,t_M$ using the QNMRF (Section \ref{sec:QNMRF}) to remove a single damped sinusoid is shown in red and shows close agreement with the pyRing single \ac{ds} model. 
}
\end{figure*}

The results from the \acp{qnmrf} pipeline, when removing a single damped sinusoid at $10 t_M$ after $t_{\rm LLO}$, are shown in Figure \ref{fig:f-tau} and are in close agreement with the pyRing single \ac{ds} results. Both methods find that the frequency and damping time inferred from the event lie between those of the (2,2,0) and (2,1,0) modes predicted by the \ac{IMR} inference.

\section{Investigation of apparent deviation from general relativity}
\label{sec:investigation}

\begin{figure}
    \centering
    \includegraphics[width=0.98\linewidth]{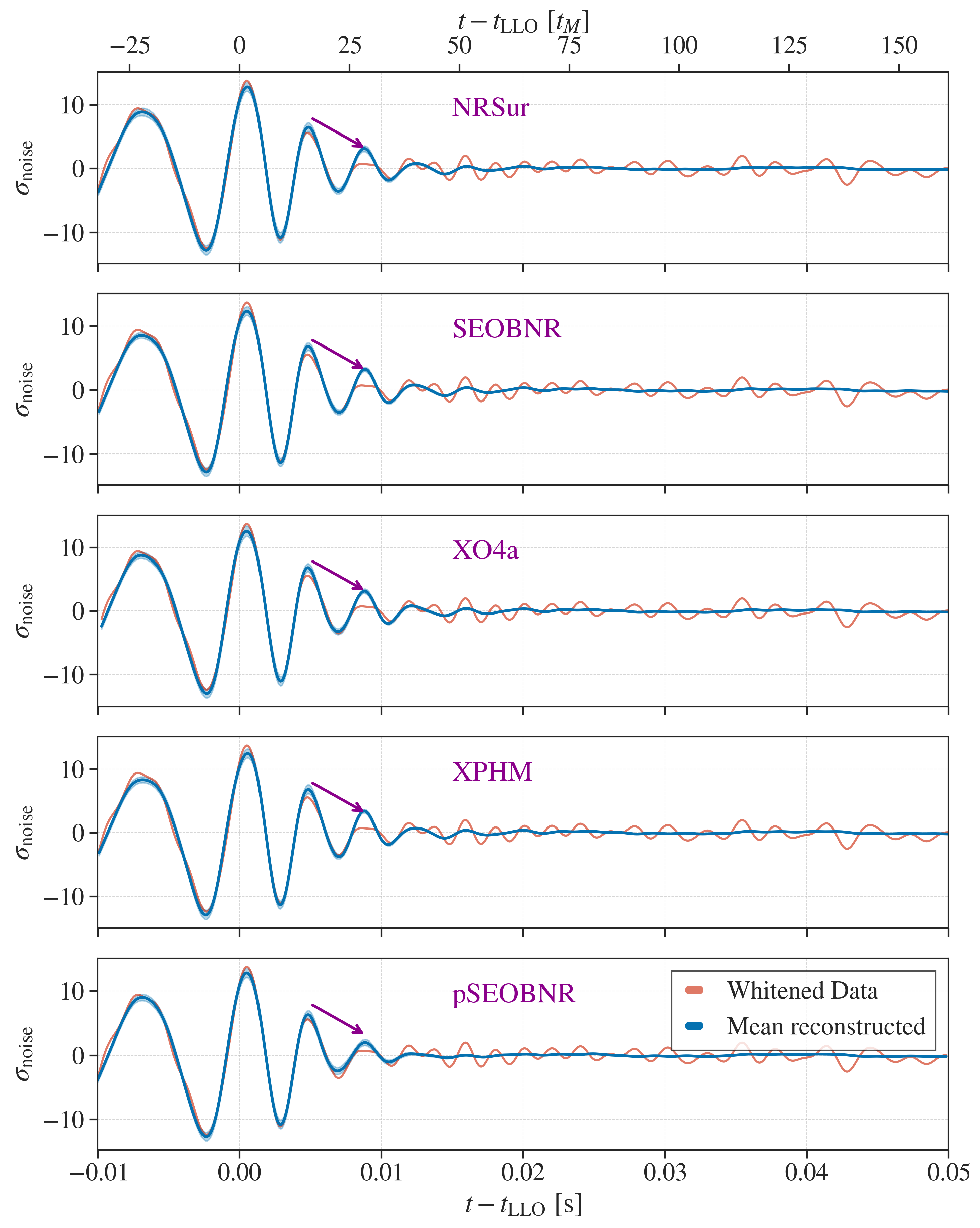}
    \caption{Comparison of whitened data and reconstructed waveforms. 
    The shaded bands around the mean encompass $90\%$ of the waveforms reconstructed from the posterior distribution. The first four panels show the four named \ac{GR} waveforms, whereas
    the fifth shows the non-GR \ac{pseobnr} waveform.
    The additional freedom in the \ac{pseobnr} waveform enables it to better capture features in the data compared
    to the four \ac{GR} waveforms, most notably by producing a quicker than expected amplitude decay. This is particularly notable at the time indicated by the arrows, \(t-t_\mathrm{LLO} \approx 0.085\,\mathrm{s}\), where the \ac{GR} waveforms all deviate noticeably from the data and consistently predict a higher than observed strain. This feature may be explained by a random fluctuation in the detector noise;
    similar fluctuations can be observed at later times (e.g., after \( t - t_\mathrm{\rm LLO} = 0.04\,\mathrm{s} \), when no remaining detectable signal is expected).}
    \label{fig:waveform_recovery_comp}
\end{figure}

The results presented in the previous section indicate that the inspiral phase of 
GW230814 is consistent with the predictions of \ac{GR}.
In contrast, the \ac{pseobnr} and TIGER analyses find that deviation parameters that affect the merger-ringdown portion of the signal have posterior probability
distributions that are peaked away from the \ac{GR} value of zero.
A non-zero value for a \ac{GR} deviation parameter does not necessarily imply a violation of \ac{GR} has been detected.
Such deviations have been observed in tests before~\citep{2021PhRvD.103l2002A,2025PhRvD.112h4080A,Maggio:2022hre} and can be caused by a variety of known effects~\citep{Gupta:2024gun}.

\begin{figure}
    \includegraphics[width=0.98\linewidth]{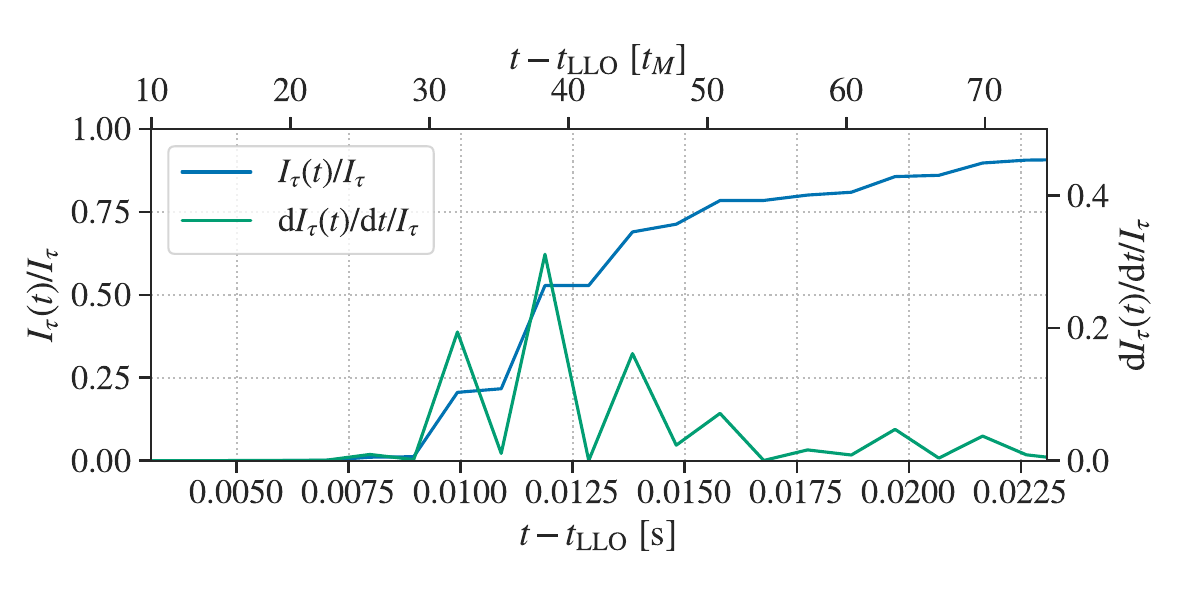}
    \caption{\label{fig:ds-fisher} The accumulation of Fisher information about the $\delta\hat{\tau}_{220}$ parameter over
    time (partial sums in blue and terms in the sum in green from the final term in Eq.~\ref{eq:fisher_tau}) for a \ac{ds} model with $f = 250 \,
    \mathrm{Hz}$ and $\tau = 3.5 \, \mathrm{ms}$, starting at $t-t_{\rm LLO} = 10 t_{\rm M}$.
    Each peak of the green curve corresponds to a peak or trough of the waveform shown in Figure~\ref{fig:waveform_recovery_comp}, with the highest peak
    corresponding to the cycle of the signal indicated by the purple arrow in that figure. A \ac{ds} analysis in the time domain is thus particularly sensitive in its damping time estimate to the times that seem to drive the deviation identified in the pSEOBNR analysis at $t-t_{\rm LLO}=9\,\mathrm{ms}$.
    }
\end{figure}

\begin{figure}
    \centering
    \includegraphics[width=0.98\linewidth]{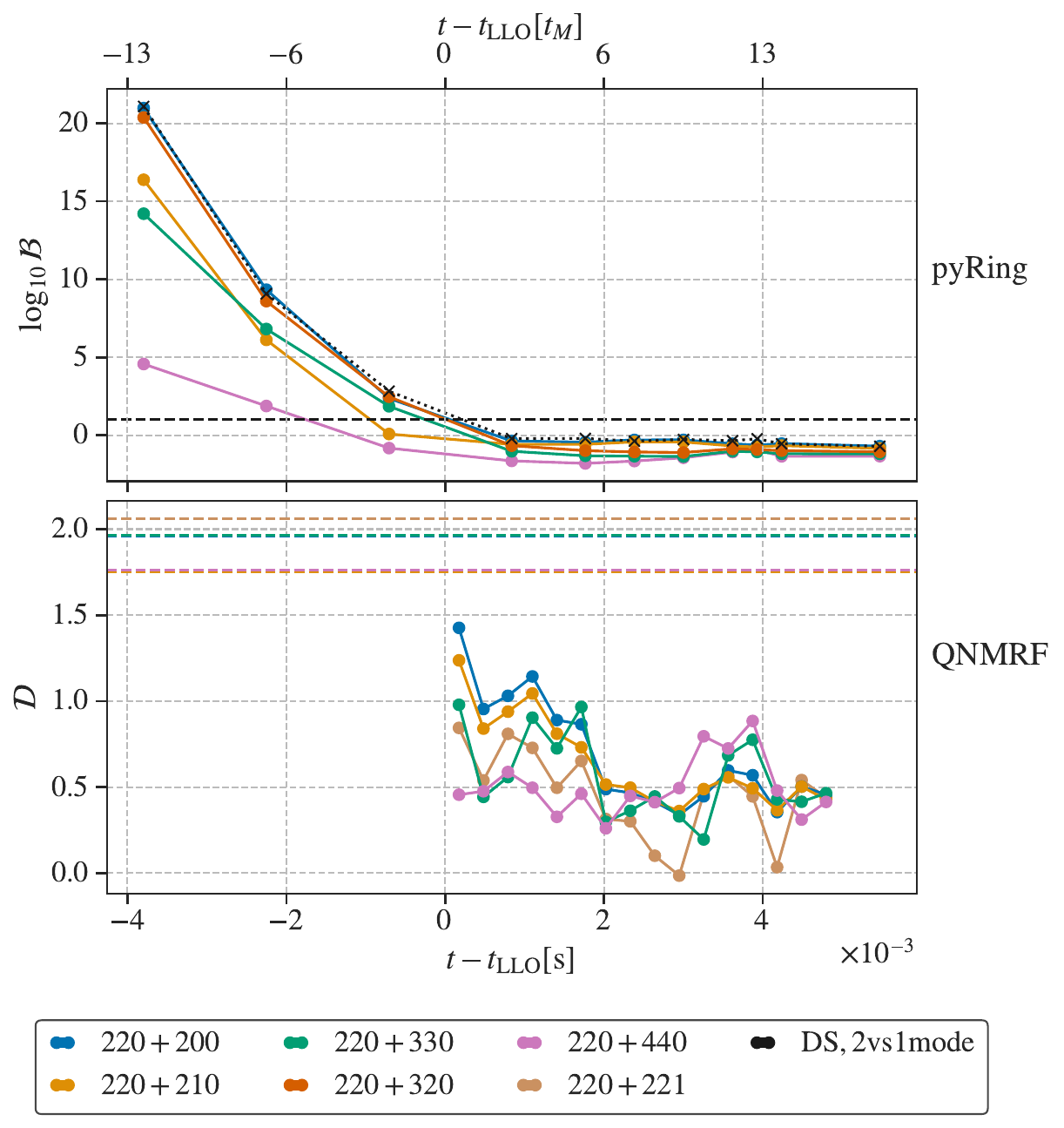}
    \caption{The $\log_{10}$ Bayes factors ($\log_{10} \mathcal B$) and detection statistics ($\mathcal{D}$) 
    from the Kerr and \ac{ds} pyRing pipelines (top) and the \ac{qnmrf} pipeline (bottom), 
    quantifying the consistency of the data with the presence of a secondary mode at various starting times.
    None of the pyRing analyses provide evidence for a secondary mode, as all yield $\mathcal{B} < 10$ (dashed black line)
    for times beyond $t_{\rm \rm LLO}$.
    Similarly, \ac{qnmrf} does not identify any two-mode combination beyond 220 that meets a FAP threshold 
    of 1\% (dashed colored lines).
    }
    \label{fig:QNMRF_result}
\end{figure}

A natural starting point to investigate the origin of these potential deviations
is to examine the data directly, by comparing the whitened strain with 
the predictions of the various waveform models employed in the analyses.
Figure~\ref{fig:waveform_recovery_comp} shows the whitened waveform posteriors
for GW230814 derived using the \ac{nrsur}, \ac{xo4a}, \ac{xphm}, \ac{seobnr} and \ac{pseobnr} models 
along with the whitened time domain data. 
While the four \ac{GR} models yield mutually consistent reconstructions, 
the \ac{pseobnr} model provides a visibly better fit to the data, 
particularly near \reviewed{$t-t_\mathrm{\rm LLO}=9$\,ms}.
At this time, the amplitude of the data appears to be lower
than predicted by \ac{GR}-based models.
In parameterized frameworks, such a discrepancy would manifest as a reduced
damping time $\tau$, a trend that is indeed observed in both the \ac{pseobnr}
and pyRing ({KerrPostmerger}) analyses.

We can confirm this by investigating the contribution to the $\delta\hat{\tau}_{220}$ inference as a function of time by evaluating the integrand
of the Fisher information matrix element in the time domain. This is shown in
Figure~\ref{fig:ds-fisher} for an analysis starting at $t- t_{\rm LLO} = 10\,t_M$.
The Fisher
information is 
\begin{equation}
    \label{eq:fisher_tau}
    I_{\tau} = \sum_{ij} \frac{\partial h_i}{\partial \tau} C^{-1}_{ij} \frac{\partial h_j}{\partial \tau} = \sum_i \left(\frac{\partial \tilde{h}_i}{\partial \tau}\right)^2,
\end{equation}
where 
\begin{equation}
    h_i = A \exp\left( -\frac{t_i}{\tau} \right) \cos\left( 2 \pi f t_i \right)
\end{equation}
is the \ac{ds} model, $C$ is the time-domain covariance matrix derived
from the estimated noise power spectral density used for the \ac{PE} 
analysis~\citep{Isi:2021iql}, and $\tilde{h} = L^{-1}
h$ with $C = L L^T$ is the whitened waveform.  For illustrative purposes, we
choose $f = 250 \, \mathrm{Hz}$ and $\tau = 3.5 \, \mathrm{ms}$, consistent with
the estimated \ac{ds} parameters from the ringdown analysis (see Figure~\ref{fig:f-tau}).  For these fiducial parameters, the Fisher
information about the parameter $\tau$ accumulates rapidly during the time
corresponding to the residual deviation at $t - t_{\rm LLO} = 9 \, \mathrm{ms}$
discussed above and shown in Figure~\ref{fig:waveform_recovery_comp}.  The
inferred damping time in a \ac{ds} ringdown analysis is thus
particularly sensitive to the data around the time of deviation shown in Figure
\ref{fig:waveform_recovery_comp}.

This observation then raises a key question: what causes this deviation?
The following sections explore whether the feature could be attributed
to additional subdominant modes in the ringdown, gravitational lensing of the signal, systematic errors in our modeling of the \ac{GR} waveform or detector noise artifacts.
While we find subdominant ringdown modes and lensing to be unlikely explanations, both waveform systematics and
detector noise could be responsible for the observed deviation.

\subsection{Tests for subdominant ringdown modes}

To test for additional subdominant ringdown modes we apply the \acf{qnmrf} and pyRing independently and compare their results.

To determine the modes present in a ringdown signal, the \acf{qnmrf} defines a detection statistic
$\mathcal{D}$ that quantifies how well the data supports a given two-mode model over a single-mode model~\cite{Lu:2025mwp}.
In this analysis,  we focus on the detection of subdominant ringdown modes by comparing a (2,2,0)-only
\ac{qnm} model to models with a different secondary mode. Specifically, we test for the presence of the (2,2,1), (2,1,0), (2,0,0), (3,2,0), (3,3,0), and (4,4,0) modes. To evaluate the detection statistic, we compute a threshold on $\mathcal{D}$ which corresponds to a statistical significance with a 1\% false-alarm probability (FAP). 
We create 300 injections by adding a single \ac{qnm} with random properties into the detector noise before and after the trigger time. 
We then compare the detection statistic of this injection between a two-mode hypothesis (for a specific secondary mode) which is known to be incorrect, 
and the correct single-mode hypothesis. We find a distribution
of detection statistics under the null hypothesis and determine the threshold on $\mathcal{D}$ which corresponds to a 1\% FAP.  This threshold is then compared to the detection statistic computed from the event which is evaluated at multiple start times $t-t_{\rm LLO}$ since the start time of the stationary ringdown regime is uncertain. The results of the \ac{qnmrf} analysis are shown in Figure~\ref{fig:QNMRF_result}. 
No two-mode hypothesis is detected with a detection statistic larger than its corresponding threshold at any analysis
start time. 

The pyRing pipeline also compares the evidence of models with a single \ac{ds} or with two \acp{ds}. The results of model selection for the presence of selected secondary modes vs the (2,2,0) mode for the KerrBH model, 
and for the uninformed \ac{ds} model, are shown in Figure~\ref{fig:QNMRF_result}. 
The Bayesian evidence slightly disfavors models with two \ac{ds} in the data rather than one with 
a $\log_{10} \mathcal{B}$ \reviewed{$\sim-0.4$}, with the KerrBH model finding similar results for its 
most probable secondary modes. This is consistent with the result found by \ac{qnmrf};
the data do not require the presence of a second ringdown mode.

\subsection{Tests for gravitational lensing}

Distortions of the \ac{GW} signal caused by gravitational lensing can masquerade as violations of \ac{GR} and they must be separately ruled out as a possible cause of apparent
\ac{GR} deviations~\citep{Gupta:2024gun,Mishra:2023vzo}.
To test for single-image lensing, investigations were performed on GW230814 using two pipelines: GOLUM~\citep{Janquart:2021qov} and Gravelamps~\citep{2022ApJ...935...68W,2023MNRAS.525.4149L}.

The GOLUM pipeline searches for strong-lensing signatures caused by distortions in the waveform from high-mass lenses~\citep{Janquart:2021qov}. This type of lens causes the formation of multiple images of the same GW, with a magnification, a time delay, and a phase shift $n_j$ of either 0, 0.5, or 1.0 (type-I, II, and III images, respectively)~\citep{2021PhRvD.103f4047E}. For single GWs such as GW230814, the pipeline searches for phase shifts of $n_j = 0.5$, which can indicate lensing~\citep{Janquart:2021nus,Wright:2024mco}.

The GOLUM pipeline's search for type-II images retrieves Bayes factors of \reviewed{$\log_{10} \mathcal{B}^{\rm II}_{\rm I} = -1.2$ and $\log_{10} \mathcal{B}^{\rm II}_{\rm III} = -1.1$} indicating that a type-II image is disfavored. Since the expectation for standard waveforms is a $\log_{10} \mathcal{B}$ of zero~\citep{Vijaykumar:2022dlp}, the disfavoring of a type-II image could be a result of either higher-order modes or other alternative deviations in the waveform that break the phase degeneracy~\citep{Wright:2024mco}.

The Gravelamps pipeline searches for waveform distortions in a single GW signal arising due to microlensing~\citep{2022ApJ...935...68W,2023MNRAS.525.4149L}. 
In general, a full wave-optics treatment is needed, analysed using a physical modelling approach with specific associated lens models. 
This manifests as a single signal with frequency-dependent modulations. In some cases, geometric-optics approximation can be applied. 
Here, the multiple images overlap and their effect manifests as a single signal with a beating pattern. 
The physical modelled search with an isolated point-mass yielded a \reviewed{$\log_{10} \mathcal{B}$ of $-0.6$}.
Thus, this signal is not considered as a candidate microlensed event.

\bigskip
\bigskip

\subsection{Assessing waveform systematics}

Waveform systematics arise from limitations in our \ac{GR} models, 
and may cause apparent deviations from \ac{GR} in our tests against real signals.
Waveform systematics are particularly concerning in the case of
signals with large \acp{SNR}, as in GW230814.
While \ac{GR}-consistent waveform models generally produce similar 
posterior distributions (see Section~\ref{sec:basic_properties}), subtle differences in the waveforms
can become significant when additional degrees of freedom are introduced, potentially mimicking 
violations of \ac{GR}~\citep{Gupta:2024gun}.

\subsubsection{\Ac{nr} simulations}

\begin{figure}
    \centering
    \includegraphics[width=0.98\linewidth]{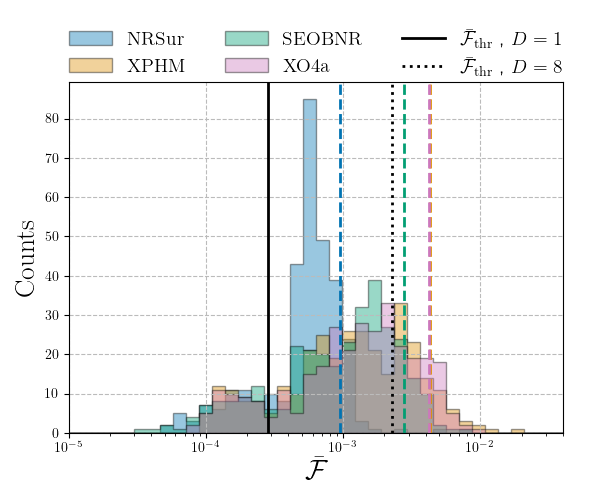}
    \caption{
        Sky-maximized, SNR-weighted unfaithfulness of the \ac{nrsur}, \ac{seobnr}, \ac{xphm} and \ac{xo4a} waveform models 
        with respect to 53 simulations (52 from the LVK \ac{nr}-SXS catalog and the targeted simulation produced as part of this
        work) compatible with the parameters of GW230814, 
        marginalized over six values of inclination. 
        Dashed vertical lines indicate the $90^{\rm th}$ percentile of the unfaithfulness distribution for each model (note that
        the \ac{xphm} and \ac{xo4a} lines overlap); black vertical lines denote the faithfulness thresholds for $D=8$
        (dotted) and $D=1$ (solid).
        The \ac{nrsur} model is the most faithful, with $>99\%$ of the simulations
        having unfaithfulness below the threshold of \reviewed{$2.3 \times 10^{-3}$}, corresponding
        to $D=8$ at the \ac{SNR} of GW230814.
    }
    \label{fig:faithfulness}
\end{figure}

Assuming that \ac{GR} signals are well captured by \ac{nr} simulations,
we assess the faithfulness of waveform models against \ac{nr} data 
within the relevant parameter space for GW230814.

We run a \ac{nr} simulation of a \ac{BBH} system 
with parameters similar to the maximum-likelihood parameters 
of the sampled posterior produced by the PE carried out using 
the \ac{nrsur} waveform model. The simulation is performed using 
the Spectral Einstein Code (SpEC) developed by 
the Simulating eXtreme Spacetimes (SXS) collaboration 
\citep{2019CQGra..36s5006B, Scheel:2025jct, SpECwebsite}, that employs a 
multi-domain spectral discretization scheme. 
The reference eccentricity of the initial data is reduced to $\sim 5\times10^{-4}$. The waveform is post-processed, 
wherein it is extrapolated to infinity using \textit{scri} 
\citep{scri} and corrected for the centre-of-mass drift 
\citep{Woodford:2019com}. The resulting waveform has about 
$42$ inspiral cycles. 
We simulate at three different resolutions to test the convergence, 
finding that at least all modes up to $\ell=4$ exhibit convergence 
of root-mean-square deviations between neighboring resolutions. 
Furthermore, we project the NR waveforms at different resolutions onto
the detector with extrinsic parameters close to the maximum likelihood
sample of the \ac{nrsur} analysis, and find the overlap (without a phase alignment)
between the the two highest resolutions to be \reviewed{$>99.998\%$}.

The accuracy requirements of waveform models
can be formulated in terms of the faithfulness $\mathcal{F}$ of the models to the true
signal~\citep{Lindblom:2008cm, Damour:2010zb, 2017PhRvD..95j4004C, Toubiana:2024car}.
Following ~\citet{2017PhRvD..95j4004C}, the faithfulness threshold $\mathcal{F}_{\rm thr}$ above which
waveform systematics do not affect the results of an analysis is
\reviewed{$\mathcal{F}_{\rm thr} = 1 - D/(2\rho^2)$}, or equivalently,
\reviewed{$\bar{\mathcal{F}}_{\rm thr} = D/(2\rho^2)$} 
where $\bar{\mathcal{F}} = 1 - \mathcal{F}$ is the unfaithfulness and 
$D$ is the number of intrinsic parameters of the source and $\rho$ is the optimal \ac{SNR}.
For GW230814, \reviewed{$D = 8$} (the two masses and 6 spin components of the two \acp{BH}) and
\reviewed{$\rho \approx 42$}, yielding \reviewed{$\bar{\mathcal{F}}_{\rm thr} \approx 2.3 \times 10^{-3}$}.
Alternatively, a more conservative threshold can be computed assuming $D=1$~\citep{Damour:2010zb}, i.e.
\reviewed{$\bar{\mathcal{F}}_{\rm thr} \approx 3 \times 10^{-4}$}.
We consider six inclination angles $\iota \in (0, \pi/2)$ and compute the sky-maximized, 
\ac{SNR}-weighted unfaithfulness~\citep{2021PhRvD.103j4056P, Harry:2017weg},
averaging over four effective-polarization angles
and four reference phases. Similar to GW231123~\citep{2025ApJ...993L..25A},
the unfaithfulness is further minimized over an in-plane rotation of the component spins.

Using this statistic, we compare the targeted \ac{nr} waveform to the
\ac{nrsur}, \ac{seobnr}, \ac{xphm} and \ac{xo4a} models. 
We find $\bar{\mathcal{F}}$ to be \reviewed{$8.1 \times 10^{-4}$, $2.2\times 10^{-3}$, $2 \times 10^{-3}$ and $2.6 \times 10^{-3}$} respectively,
averaging over the six inclination angles.
Expanding the investigation to nearby areas of the parameter space, from approximately 1700 SXS
simulations in the LVK \ac{nr} catalog we identify \reviewed{52}
that lie within the \reviewed{$90\%$} credible ranges of mass ratio $q$, effective inspiral spin $\chi_{\rm eff}$, 
and precession parameter $\chi_p$ for this event.
These results are shown in Figure~\ref{fig:faithfulness}.
We find that in this region of the parameter space the \ac{nrsur} model is the most accurate with respect to the
simulations considered, followed by the \ac{seobnr}, \ac{xphm}, and \ac{xo4a} models. 
With respect to the accuracy requirements considered, the \ac{nrsur} model satisfies the $D=8$
threshold for \reviewed{more than $99\%$ of the systems}, and the $D=1$ threshold for
\reviewed{$\sim 17\%$} of the systems. 
Overall, these results indicate that \reviewed{waveform systematics could be a contributing factor for the apparent \ac{GR} deviation.}

\bigskip

\subsubsection{Zero-noise waveform injections}
 
\begin{figure}[htbp]
    \centering
    \includegraphics[width=0.98\linewidth]{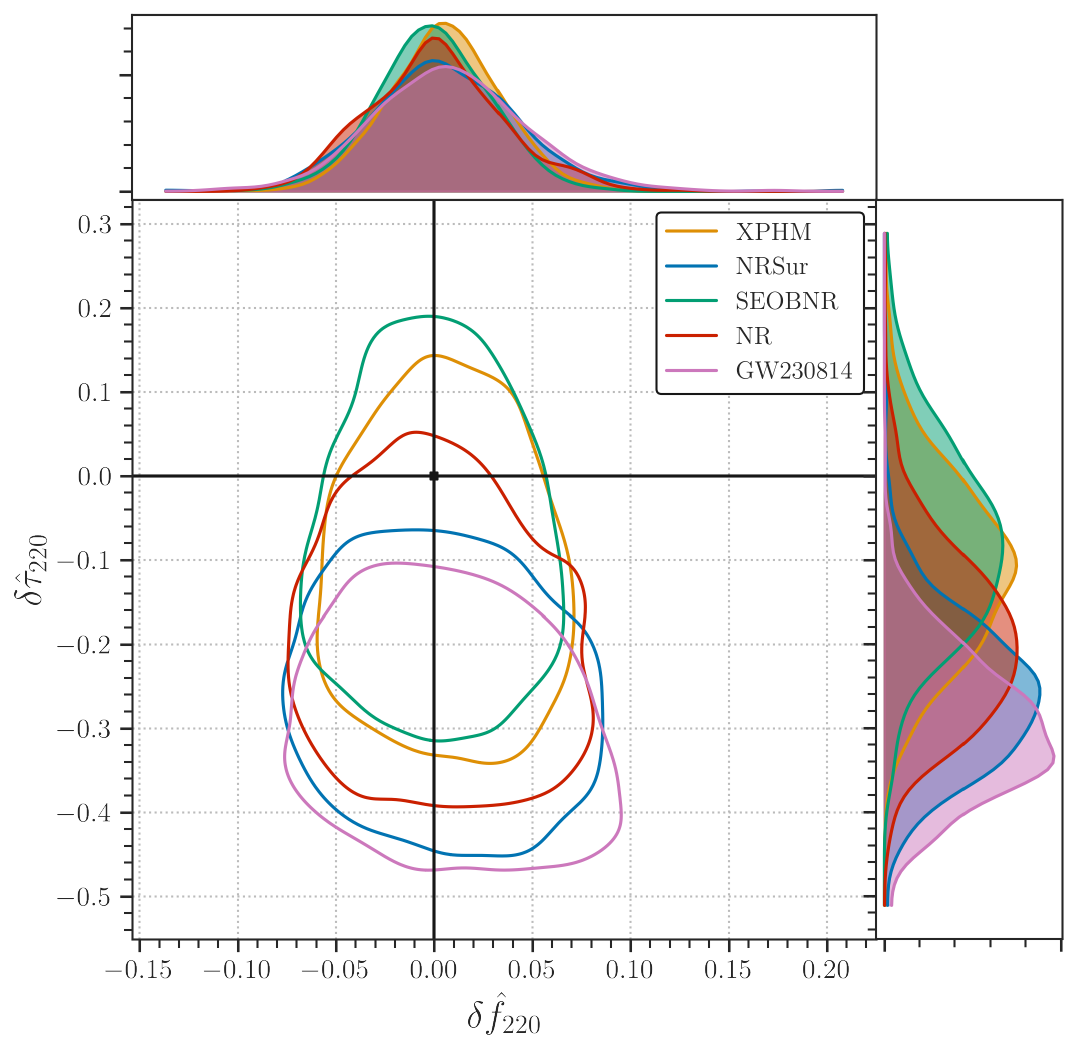}
    \caption{Comparison of the \ac{pseobnr} analyses on injections in zero noise of maximum-likelihood
    waveforms from different waveform models. The \ac{seobnr} and \ac{xphm} injections are recovered consistent with \ac{GR}. 
    The \ac{nrsur} and \ac{nr} injections show a deviation more similar to the one found for the actual signal (in pink).}
    \label{fig:pSEOB_zeronoise}
\end{figure}

The accuracy thresholds discussed above are sufficient, not necessary, conditions.
A model falling short of the threshold may still yield unbiased results for the source parameters, as is observed in the standard 
\ac{seobnr} and \ac{xphm} analyses of Section~\ref{sec:basic_properties}. Furthermore, such criteria provide no information on which parameters
are most susceptible to potential biases, and are known to be overly conservative~\citep{2020PhRvR...2b3151P, Toubiana:2024car, Finch:2021iip}.

To further investigate the impact of waveform systematics, 
we perform zero-noise injections using the maximum-likelihood waveforms from
the \ac{nrsur}, \ac{seobnr}, and \ac{xphm} models, and the \ac{nr} injection, and recover them using the \ac{pseobnr} waveform model.
Figure~\ref{fig:pSEOB_zeronoise} presents these results.

The \ac{seobnr} and \ac{xphm} injections
yield results consistent with \ac{GR}. However, we observe that the \ac{GR} value lies at the edge of the $90\%$
credible interval of the $\delta \hat{\tau}_{220}$ posterior for both the \ac{nr} and \ac{nrsur} injections, 
a shift resembling that observed in the real data.

To test the robustness of this effect, we conduct \reviewed{10} additional zero-noise injections using 
\ac{nrsur} waveforms drawn from the \ac{PE} samples of the event, again analyzing them with \ac{pseobnr}.
\reviewed{Two} of the ten injections show a deviation in the damping-time similar to the GW230814 analysis.
The injected parameters are recovered with good accuracy except for the spin magnitudes
and tilt angles.

This lends support to the hypothesis that waveform systematics 
could be contributing to the observed biased recovery of the ringdown damping-time by the \ac{pseobnr} analysis.

\subsection{Assessing detector effects}

An alternative explanation for the apparent deviations is that they may be
caused by fluctuations of the detector noise. No data-quality issues were reported
at the time of GW230814, but low-amplitude noise fluctuations may still be present in the data.
Noise fluctuations may be broadly classified as fluctuations consistent with
a Gaussian distribution, or as low-amplitude non-Gaussian noise that does 
not pass the threshold for glitch identification~\citep{2025CQGra..42h5016S}.
For example, Figure~\ref{fig:waveform_recovery_comp} shows that
noise fluctuations of approximately the same amplitude as the feature previously highlighted in the whitened data
are also present at later times, after the black hole has settled into a stationary state
and the signal has died away.
To investigate the impact of noise, we inject GR-waveforms
in Gaussian and real noise, and analyse them
with our pipelines to determine how often we
find results similarly extreme to those found for GW230814.

\subsubsection{\ac{pseobnr} injection results}
For the \ac{pseobnr} pipeline, the analysis of GW230814 is compared with: 
(i) ten \ac{nrsur} injections in zero noise with parameters taken from the
analysis of the event with \ac{nrsur}; 
(ii) ten \ac{seobnr} injections in Gaussian noise with the maximum-likelihood values
of the \ac{PE} of the event with \ac{seobnr}; 
(iii) twenty \ac{seobnr} injections in real \ac{LLO} noise with the maximum-likelihood values 
from the \ac{PE} of the event with \ac{seobnr}; 
(iv) ten \ac{nrsur} injections in real \ac{LLO} noise with the maximum-likelihood values 
from the \ac{PE} of the event with \ac{nrsur}.

\begin{figure}
    \centering
    \includegraphics[width=\linewidth]{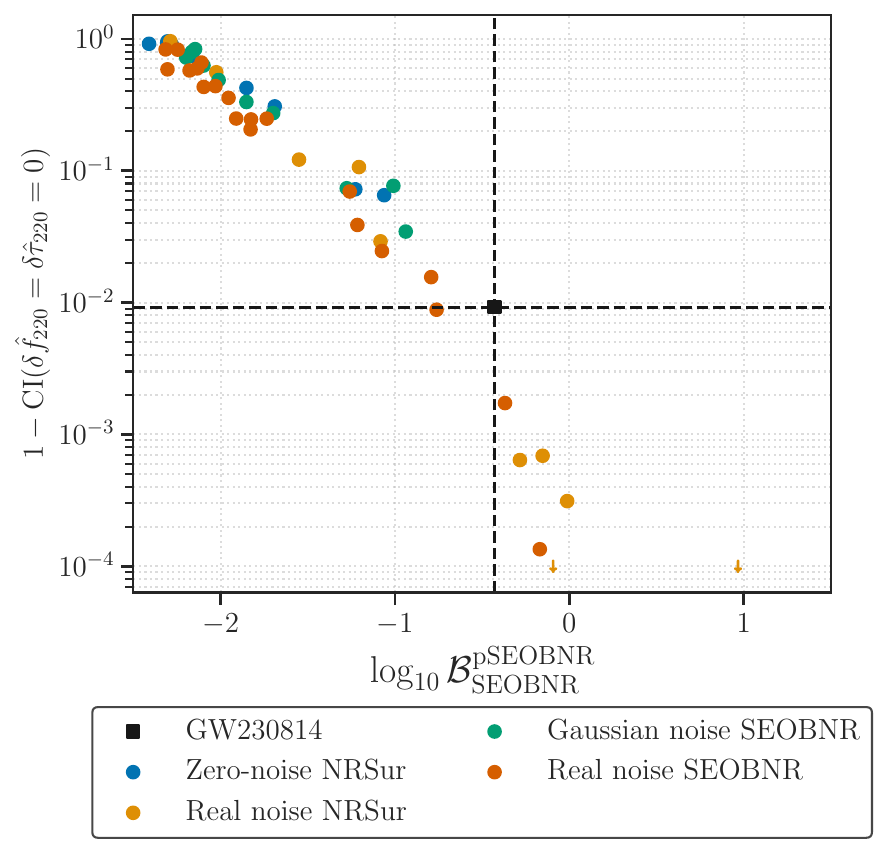}
    \caption{
        Distribution of $\log_{10} \mathcal{B}^{\rm pSEOBNR}_{\rm SEOBNR}$ versus $1 - \mathrm{CI}(\delta \hat{f}_{220}=\delta \hat{\tau}_{220}=0)$, 
        the posterior exclusion of the \ac{GR}-consistent values of damping time and frequency from the \ac{pseobnr} pipeline. Shown are results from zero-noise \ac{nrsur} 
        injections (blue), Gaussian-noise \ac{seobnr} injections (green), and real-noise injections using either \ac{seobnr} (red) or \ac{nrsur} (yellow).
        Gaussian noise or waveform systematics alone do not yield Bayes factors as large as GW230814's. 
        In contrast, real noise, especially when combined with systematics, can reproduce or exceed GW230814's Bayes factor in \reviewed{50\%} of the cases.
    }
    \label{fig:injs}
\end{figure}

The results are shown in Figure~\ref{fig:injs}.
We find that \reviewed{five} real-noise \ac{nrsur} injections and \reviewed{two} real-noise \ac{seobnr} 
injection display a Bayes factor higher than one of GW230814. For these injections, the posterior quantiles at which \ac{GR} 
is found is \reviewed{lower} than the one for GW230814. 
These results indicate that such deviations can occur due to statistical fluctuations in real detector noise.

\subsubsection{pyRing injection results}
The pyRing pipeline was used to analyse 198 injections of the maximum-likelihood \ac{nrsur} waveform into the real noise of the \ac{LLO} detector
at various times around the event, using the KerrPostmerger model of the ringdown with $\delta f$ and $\delta \tau$ parameterised deviations of the (2,2,0) mode enabled.
The evidence is compared to that of the same model with no deviations enabled to produce a Bayes factor between this non-GR model and the equivalent GR one. 
We find a Bayes factor that is greater than that for the observed signal in \reviewed{59\%} of cases. Using the same set of simulations, we find credible intervals that are more extreme than those measured for the observed signal in \reviewed{32\%}
of cases for the damping time and \reviewed{20\%} of cases for the frequency.
We conclude that the results of the pyRing analysis are consistent with the distribution under noise for similar signals.

\section{Conclusions}\label{sec:conclusion}
GW230814, observed exclusively by the LIGO Livingston detector, exhibits the highest \ac{SNR} among the signals of \ac{gwtc4}~\citep{LIGOScientific:2025slb}, which enables tight constraints on its intrinsic parameters. We identify its source as a \ac{BBH} coalescence, with component masses and spins consistent with the known population of \ac{BBH} mergers.
The strong higher-order modes present in the waveform allow for constraints on the inclination angle, and together with the high \ac{SNR} enable precise measurement of the mass ratio, despite the absence of a second detector to resolve the full polarization content. These measurements yield well-constrained estimates of the component masses.

Tests of the PN coefficients of the inspiral show consistency with  \ac{GR}, with several constraints from this single signal comparable to population constraints from the whole of \ac{gwtc3}~\cite{2025PhRvD.112h4080A}.
However, as shown in Figures~\ref{fig:tiger_result},~\ref{fig:pSEOB_result} and~\ref{fig:f-tau} the TIGER, \ac{pseobnr} and pyRing-KerrPostmerger tests reveal mild inconsistencies in the post-inspiral portion of the signal.
These deviations are consistent across methods and suggest a reduction in the damping time of the least-damped quasi-normal mode.
Figure~\ref{fig:waveform_recovery_comp} shows a whitened data segment of the ringdown exhibiting faster amplitude decay than predicted by standard \ac{GR} templates.
Given the limited contribution of this segment to the overall waveform, its impact on GR parameter estimation is expected to be minimal.

We explored several plausible causes of the observed deviations, including subdominant ringdown modes, gravitational lensing, waveform systematics, and detector noise fluctuations.
No compelling evidence was found for subdominant modes or lensing effects.
We show that similar inconsistencies as observed for GW230814 can be generated for simulated, \ac{GR} based signals in zero noise and added to both Gaussian noise and, more frequently, real detector noise. 
Comparable deviations also emerged when analyzing signals with mismatched waveform models, including numerical relativity injections at similarly high \ac{SNR}.

Most importantly, we cannot exclude statistical noise fluctuations as the cause of these deviations. The lack of data from a second detector severely limits our ability to distinguish between weak non-Gaussian features in the gravitational wave signal and statistical fluctuations in the noise. This underscores the need for a global network of gravitational-wave observatories for robust tests of fundamental physics.

Strain data from the LIGO detectors associated with GW230814 are available from the \citet{GWOSC}. Samples from posterior distributions of the source parameters, additional materials, and notebooks for reproducing the figures are available on Zenodo \citep{zenodo}.

\section*{Acknowledgments}
This material is based upon work supported by NSF's LIGO Laboratory, which is a
major facility fully funded by the National Science Foundation.
The authors also gratefully acknowledge the support of
the Science and Technology Facilities Council (STFC) of the
United Kingdom, the Max-Planck-Society (MPS), and the State of
Niedersachsen/Germany for support of the construction of Advanced LIGO 
and construction and operation of the GEO\,600 detector. 
Additional support for Advanced LIGO was provided by the Australian Research Council.
The authors gratefully acknowledge the Italian Istituto Nazionale di Fisica Nucleare (INFN),  
the French Centre National de la Recherche Scientifique (CNRS) and
the Netherlands Organization for Scientific Research (NWO)
for the construction and operation of the Virgo detector
and the creation and support  of the EGO consortium. 
The authors also gratefully acknowledge research support from these agencies as well as by 
the Council of Scientific and Industrial Research of India, 
the Department of Science and Technology, India,
the Science \& Engineering Research Board (SERB), India,
the Ministry of Human Resource Development, India,
the Spanish Agencia Estatal de Investigaci\'on (AEI),
the Spanish Ministerio de Ciencia, Innovaci\'on y Universidades,
the European Union NextGenerationEU/PRTR (PRTR-C17.I1),
the ICSC - CentroNazionale di Ricerca in High Performance Computing, Big Data
and Quantum Computing, funded by the European Union NextGenerationEU,
the Comunitat Auton\`oma de les Illes Balears through the Conselleria d'Educaci\'o i Universitats,
the Conselleria d'Innovaci\'o, Universitats, Ci\`encia i Societat Digital de la Generalitat Valenciana and
the CERCA Programme Generalitat de Catalunya, Spain,
the Polish National Agency for Academic Exchange,
the National Science Centre of Poland and the European Union - European Regional
Development Fund;
the Foundation for Polish Science (FNP),
the Polish Ministry of Science and Higher Education,
the Swiss National Science Foundation (SNSF),
the Russian Science Foundation,
the European Commission,
the European Social Funds (ESF),
the European Regional Development Funds (ERDF),
the Royal Society, 
the Scottish Funding Council, 
the Scottish Universities Physics Alliance, 
the Hungarian Scientific Research Fund (OTKA),
the French Lyon Institute of Origins (LIO),
the Belgian Fonds de la Recherche Scientifique (FRS-FNRS), 
Actions de Recherche Concert\'ees (ARC) and
Fonds Wetenschappelijk Onderzoek - Vlaanderen (FWO), Belgium,
the Paris \^{I}le-de-France Region, 
the National Research, Development and Innovation Office of Hungary (NKFIH), 
the National Research Foundation of Korea,
the Natural Sciences and Engineering Research Council of Canada (NSERC),
the Canadian Foundation for Innovation (CFI),
the Brazilian Ministry of Science, Technology, and Innovations,
the International Center for Theoretical Physics South American Institute for Fundamental Research (ICTP-SAIFR), 
the Research Grants Council of Hong Kong,
the National Natural Science Foundation of China (NSFC),
the Israel Science Foundation (ISF),
the US-Israel Binational Science Fund (BSF),
the Leverhulme Trust, 
the Research Corporation,
the National Science and Technology Council (NSTC), Taiwan,
the United States Department of Energy,
and
the Kavli Foundation.
The authors gratefully acknowledge the support of the NSF, STFC, INFN and CNRS for provision of computational resources.

This work was supported by MEXT,
the JSPS Leading-edge Research Infrastructure Program,
JSPS Grant-in-Aid for Specially Promoted Research 26000005,
JSPS Grant-in-Aid for Scientific Research on Innovative Areas 2402: 24103006,
24103005, and 2905: JP17H06358, JP17H06361 and JP17H06364,
JSPS Core-to-Core Program A.\ Advanced Research Networks,
JSPS Grants-in-Aid for Scientific Research (S) 17H06133 and 20H05639,
JSPS Grant-in-Aid for Transformative Research Areas (A) 20A203: JP20H05854,
the joint research program of the Institute for Cosmic Ray Research,
University of Tokyo,
the National Research Foundation (NRF),
the Computing Infrastructure Project of the Global Science experimental Data hub
Center (GSDC) at KISTI,
the Korea Astronomy and Space Science Institute (KASI),
the Ministry of Science and ICT (MSIT) in Korea,
Academia Sinica (AS),
the AS Grid Center (ASGC) and the National Science and Technology Council (NSTC)
in Taiwan under grants including the Science Vanguard Research Program,
the Advanced Technology Center (ATC) of NAOJ,
and the Mechanical Engineering Center of KEK.

Additional acknowledgements for support of individual authors may be found in the following document: \\
\url{https://dcc.ligo.org/LIGO-M2300033/public}.
For the purpose of open access, the authors have applied a Creative Commons Attribution (CC BY)
license to any Author Accepted Manuscript version arising.
We request that citations to this article use `A. G. Abac {\it et al.} (LIGO-Virgo-KAGRA Collaboration), ...' or similar phrasing, depending on journal convention.

The \ac{nr} simulations were run at the \textit{sonic} High Performance Computing facility at ICTS-TIFR, Bengaluru, India 
supported by the Department of Atomic Energy, Government of India, under Project No. RTI4001.

\software{
The following software was used in this research:
Asimov \citep{2023JOSS....8.4170W};
Astropy \citep{astropy:2013, astropy:2018, astropy:2022};
BayesWave \citep{bayeswave};
Bilby \citep{2019ApJS..241...27A,2020MNRAS.499.3295R,2020MNRAS.498.4492S};
Bilby-TGR \citep{ashton_2025_15676285};
CPNest \citep{cpnest};
Dynesty \citep{dynesty};
FTI \citep{Mehta:2022pcn};
GOLUM \citep{Janquart:2021qov};
GraceDB \citep{gracedb};
Gravelamps \citep{2022ApJ...935...68W,2023MNRAS.525.4149L};
GSTLAL \citep{2017PhRvD..95d2001M, 2019arXiv190108580S, 2020PhRvD.101b2003H, 2021SoftX..1400680C, sakon2023template, ewing2023performance, 2023PhRvD.108d3004T};
GWpy \citep{gwpy};
LALSuite \citep{lalsuite,swiglal};
MBTA \citep{2021CQGra..38i5004A};
PESummary \citep{pesummary};
PyCBC \citep{pycbc};
PyRing \citep{Carullo:2019flw,pyRing};
PySEOBNR \citep{Mihaylov:2023bkc};
\ac{qnmrf} \citep{Ma:2022wpv,Ma:2023vvr,Ma:2023cwe};
Ringdown \citep{Isi:2021iql};
Spec \citep{SpECwebsite2};
TIGER \citep{AgathosEtAl:2014, 2018PhRvD..97d4033M, Roy:2025gzv}
}.

Late in the preparation of this manuscript, an error was discovered in the likelihood evaluation within Bilby (see \citet{Talbot:2025vth} and \cite{LIGOScientific:2025zdk} for details). Consequently, we repeated all GR PE runs and the \ac{pseobnr} analysis using the corrected likelihood. A complete rerun of the simulated signals (including \ac{pseobnr}) and the FTI, TIGER, PCA, and MDR tests was not feasible. However, since the reruns validated that a likelihood reweighting would produce consistent results, a complete rerun was deemed unnecessary. Therefore, we have updated the results from these specific analyses via reweighting, which appropriately corrects the posterior distributions, log-likelihoods, and Bayes factors.

\bibliographystyle{aasjournal}
\bibliography{gwtc-common-files/references/bibliography,gwtc-common-files/references/compendia,references}

\end{document}